\documentclass[sigconf,nonacm,table,xcdraw]{acmart}

\usepackage[colorinlistoftodos]{todonotes}
\usepackage{comment}
\usepackage{microtype}
\usepackage{wrapfig}
\usepackage{algorithm}
\usepackage[noend]{algpseudocode}
\usepackage{subcaption}
\usepackage{needspace}
\usepackage{url}
\usepackage{amsmath} 
\usepackage{soul}
\usepackage{makecell}
\usepackage{multirow}
\AtBeginDocument{%
  \providecommand\BibTeX{{%
    \normalfont B\kern-0.5em{\scshape i\kern-0.25em b}\kern-0.8em\TeX}}}

\newcommand{\AD}[1]{\todo[color=cyan,inline]{Aditya: #1}}

\begin{document}
\title{LGNNIC: Acceleration of Large-Scale GNN Training using SmartNICs}

\author{Liad Gerstman}
\authornote{Part of the work was done while the author was an intern at Hewlett Packard Labs}
\affiliation{
 \institution{Technion Israel Institute of Technology}
 \country{Haifa, Israel}
}

\email{liadgerstman@campus.technion.ac.il}

\author{Aditya Dhakal}
\affiliation{
 \institution{Hewlett Packard Enterprise Labs}
 \country{Fort Collins, CO, USA}
}

\email{aditya.dhakal@hpe.com}


\author{Dejan Milojicic}
\affiliation{
 \institution{Hewlett Packard Enterprise Labs}
 \country{Milpitas, CA, USA}
}
  
\email{dejan.milojicic@hpe.com}

\author{Avi Mendelson}
\affiliation{
 \institution{Technion Israel Institute of Technology}
 \country{Haifa, Israel}
}

\email{mendlson@technion.ac.il}

\begin{abstract}
Graph Neural Networks (GNNs) are widely used across domains such as natural sciences, social network analysis, chip design, and recommendation systems. However, as graph sizes grow, storing and processing them entirely on a single-node CPU-GPU system becomes increasingly impractical. A promising approach is to distribute the graph across multiple remote memory nodes, though this introduces a major bottleneck: inter-node network congestion during training. To address this, we propose LGNNIC, a novel inter-node system architecture that leverages SmartNICs co-located with remote memory nodes---a configuration already available in modern systems---to reduce communication overhead in distributed GNN training. LGNNIC offloads key preprocessing tasks to SmartNICs, reducing the volume of data transferred to computational (training) nodes and alleviating network congestion.

We introduce two complementary techniques executed on the SmartNICs during the preprocessing phase: Neighbor Sampling, which performs mini-batch sampling, and Quantization of the sampled batches. To evaluate LGNNIC under different communication infrastructures, we designed both an optimized low-overhead DMA-based synchronization mechanism and a high-overhead socket-based alternative used as a benchmark. We evaluate the core SmartNIC offloading mechanisms across standard GNN workloads and sampling hyperparameters using a proof-of-concept (PoC) system comprising one remote-memory node with an NVIDIA BlueField-2 SmartNIC and one compute node with an A100 GPU. 
Both Neighbor Sampling and Quantization on the remote node demonstrated substantial training speedups in most configurations. Neighbor Sampling achieved up to a 62.4x speedup with Sockets and 17.5x with DOCA-DMA on commonly used hyperparameters across datasets, primarily due to reductions in data transaction time (up to 73.6x and 5.1x, respectively). Quantization contributed additional speedups of up to 3.6x with Sockets and 1.3x with DOCA-DMA, also due to reduced data transfer (up to 3.6x and 2x, respectively).
\end{abstract}
\maketitle

\section{Introduction}
Graph Neural Networks (GNNs) have attracted interest from the research community for machine-learning-based relational systems. Several classes of GNNs have been studied: Graph Convolutional Networks (GCN) \cite{ref:kipf2017semisupervised}, Graph Attention Networks (GAT) \cite{ref:velickovic2018graph}, GraphSAGE \cite{ref:hamilton2018inductive} and Graph Isomorphism Network (GIN) \cite{ref:gin}. These GNN architectures differ in how they aggregate and transform information from neighboring nodes.

GNN training produces node embeddings used for tasks such as node classification, graph classification, and link prediction. GNNs learn node embeddings by aggregating a node’s features and the features of its connected neighbors. Computing node embeddings is a core operation in GNN training and is performed recursively over each node and its neighbors.

The challenges for GNN training arise from the exponential growth of the computational graphs as the graph size grows. With each GNN layer, neighbors of nodes from previous layers are aggregated, leading to exponential growth in computation. Due to the simultaneous calculation of node embeddings for the same layer across multiple computation graphs, it is necessary to keep the full graph's data, along with intermediate data from the embedding calculation, on the GPU. For many graphs, e.g., the heterogeneous academic graph MAG240 (167 GB) \cite{ref:hu2021ogblsc} and the encyclopedic knowledge graph WikiKG90Mv2 (89 GB) \cite{ref:hu2021ogblsc}, these raw graph sizes can easily exceed high-bandwidth memory (HBM) capacity of GPUs (40-100 GB) and their intermediate training data would require more memory. Consequently, it has become impractical to perform full-batch GNN training on a single CPU-GPU system for such datasets. A common solution involves partitioning the large graph into smaller mini-batches (during preprocessing) that can fit within the GPU's memory.

As a result of graph size explosion, sampling methods, which are typically performed on the CPU, are used (during mini-batch preparation) to reduce either the size of the graph or the mini-batches, allowing them to fit into GPU memory while maintaining a reasonable level of accuracy. An example of such a method is Neighbor Sampling, as introduced in GraphSAGE \cite{ref:hamilton2018inductive}. When scaling up to even larger GNNs beyond the point where they no longer fit even in host memory, sampling on the CPU in a single CPU-GPU node becomes infeasible. Distributing the graph across multiple memory nodes could serve as a potential solution. However, this approach introduces a significant communication bottleneck between the distributed memory and compute nodes (CPU and/or GPU), since the distribution usually requires transferring large volumes of data.

\begin{table}[t]
\centering
  \caption{Attributes of the examined PyG datasets}
   \label{tab:dataset-info}
  \scalebox{1}{
  \begin{tabular}{|>{\columncolor{gray!20}}c|c|c|c|c|}
    \hline
    \rowcolor{gray!20}
    Dataset & Nodes  & Edges & Features & Classes \\
    \hline
    Reddit & 232,965 & 114,615,892 & 602 & 41 \\
    \hline
    \makecell{OGBN-\\ Products} & 2,449,029 & 61,859,140 & 100 & 47 \\
    \hline
    \makecell{OGBN-\\ MAG} & 1,939,743 & 21,111,007 & 128 & 349 \\
    \hline
  \end{tabular}
  }
\end{table}

To address the challenges of GNN learning on distributed graphs, we present LGNNIC (Large-Scale GNN Training using SmartNICs). Our proposed system consists of multiple remote memory nodes---memory and/or storage devices equipped with SmartNICs, a common configuration in modern systems---and multiple local computational nodes equipped with CPUs and GPUs.

LGNNIC aims to alleviate communication bottlenecks through two complementary data manipulation methods: Neighbor Sampling and Tensor Quantization, both executed on the SmartNICs during the preprocessing (mini-batching) phase. These methods are executed before transferring the mini-batches to the computational nodes for the GPU training phase. We use SmartNICs to create graph samples from large graphs and we further quantize the mini-batch feature tensors before sending the sampled graph over the network.

In this paper, we first present results from executing mini-batch GNN training with and without in-layer Neighbor Sampling, as introduced by GraphSAGE \cite{ref:hamilton2018inductive}, on a local CPU-GPU system. This setup provides a baseline, motivating the subsequent profiling of different training phases executed across various components of the LGNNIC system. We then compare preprocessing times with and without Neighbor Sampling on a BlueField-2 SmartNIC (referred to as BF-2 hereafter) versus a local CPU. This comparison underscores the challenges of running preprocessing on a less powerful device near memory, while demonstrating the benefits of Neighbor Sampling in significantly reducing preprocessing time. The impact is further amplified by reduced data transfer times through remote Neighbor Sampling.

Following this, we analyze end-to-end execution times for GNN training phases. Our analysis highlights the reductions in data transfer and preprocessing times on the LGNNIC proof-of-concept (PoC) system, resulting in an overall acceleration of the training process when Neighbor Sampling is applied. Our findings indicate up to an 8.7x speedup in data transfer time and up to a 17.5x speedup in total training time when using our DOCA-DMA-based \cite{ref:docadma} mechanism and greater speedups with the socket-based mechanism. 

In our quantization approach, we evaluate the effect of quantizing mini-batch feature tensors from 32-bit floating-point (FP32) to 16-bit floating-point (FP16) across all datasets and synchronization mechanisms. Quantization is performed on the SmartNIC of the remote node before transferring the tensors to the computational node. Upon arrival, the tensors are dequantized back to FP32 to continue computation. This technique reduces transfer overhead and improves training performance with small changes in test accuracy. Notably, quantization is handled entirely on the SmartNIC---not on the remote CPU---since, in our system (as in modern architectures), data flows directly from memory through the SmartNIC without requiring an additional copy that would be necessary if routed via the CPU. Quantization achieves up to a 1.3x training speedup with DOCA-DMA and up to 3.6x with the socket-based mechanism.


To summarize our key contributions:
\begin{itemize}
\item We introduce LGNNIC, a novel architecture based on the BlueField-2 SmartNIC designed to accelerate total training time by mitigating network bottlenecks, thereby enhancing the performance of large-scale GNN algorithms.
\item We conduct an in-depth analysis of the effects of Neighbor Sampling in GNN training, comparing performance on a lower-performance compute device, such as a SmartNIC, with that on a high-performance device, such as the AMD EPYC 7513.
\item We profile the execution times of key GNN phases across a baseline CPU-GPU local node and across our LGNNIC system using three well-known GNN workloads.
\item We present a comprehensive implementation of a latency-optimized synchronization mechanism that integrates PyG and DOCA-DMA between the remote and local nodes and compare it to our simple socket-based mechanism.
\item We present the impact of quantization as an additional method to reduce data transfers and accelerate total training time.
\end{itemize}

\section{Background}
In this section, we provide background on the libraries and techniques used in this work.
\subsection{PyTorch Geometric (PyG)}
\label{subsection:PyTorch Geometric (PyG) Datasets}

Popular Python libraries for training GNNs include PyG (PyTorch Geometric \cite{ref:Fey/Lenssen/2019}) and DGL (Deep Graph Library) \cite{ref:dgl}. We utilized PyG and its class \texttt{NeighborLoader} for training with mini-batch Neighbor Sampling. To establish the baseline mini-batch training sampling performance with a local node CPU-GPU system, we selected three medium-sized datasets (Reddit, OGBN-Products and OGBN-MAG). The datasets were specifically chosen because they are popular in various papers \cite{ref:hamilton2018inductive,ref:hu2021ogblsc,ref:Cluster-GCN,ref:lee2022smartsage} and because, as shown in Table~\ref{tab:dataset-info}, they exhibit varying characteristics in terms of nodes, edges, and features, allowing us to evaluate the mechanism across graphs with different characteristics. In addition, these datasets cover a range of use cases.

The Reddit dataset is used to predict the community to which a new post belongs, based on the comments on its previous post~\cite{ref:hamilton2018inductive}. The second dataset, OGBN-Products, involves predicting the category of an Amazon product based on products purchased together \cite{ref:hu2020ogb}. The third dataset, heterogeneous OGBN-MAG, focuses on predicting the conference or journal to which each paper belongs, based on its content, references, authors, and authors' affiliations \cite{ref:hu2020ogb}. It is crucial to note that the selection of these datasets was influenced by the architecture of the proof-of-concept system, which consists of a single BlueField-2 SmartNIC, one CPU, and one GPU. In our research, we deliberately chose the largest, widely used real-world graphs that could fit within the memory constraints of these devices, enabling us to effectively demonstrate the benefits of the system.

\subsection{Neighbor Sampling With PyG}
\label{subsection:Neighbor Sampling with pyg}
With PyG, during each epoch, mini-batch sampling is performed in a batch-iterative manner by calling the \texttt{next()} function on the \texttt{NeighborLoader} iterable data loader, often named \texttt{train\_loader}. The \texttt{train\_loader} is instantiated with several arguments. Two of the most important arguments are: (1) \texttt{num\_neighbors}---a list specifying the maximum number of neighbors to sample at each layer. For example, \([25, 10]\) indicates a limitation of two layers: the first with up to 25 neighbors per node and the second with up to 10 neighbors per node, with any subsequent layers being truncated. In contrast, \([-1, -1]\) also imposes a two-layer limit but without applying in-layer Neighbor Sampling. (2)
 \texttt{num\_workers}---the number of subprocesses to run in parallel across multiple CPU cores using the CPU Affinity feature.

To cover the full graph during mini-batch training, each PyG epoch iterates over the \texttt{NeighborLoader} class (a subclass of PyTorch’s \texttt{DataLoader}) to generate sampled mini-batches. For each mini-batch, the sampled nodes, features, and labels are transferred to the GPU, where training operations such as the forward pass, loss computation, and backpropagation occur. While preloading the full graph into GPU memory can reduce sampling overhead by avoiding repeated feature transfers, this is infeasible in scenarios where the GPU lacks sufficient memory for both data and intermediate computations---as is the case in our study. 

Our optimization of this workflow with DOCA-DMA is detailed in Algorithms~\ref{alg:BlueField} and~\ref{alg:host}. For completeness, we also provide the standard PyG neighbor-sampling training loop in Appendix~\ref{subsection: appendix: Neighbor Sampling With PyG}, which serves as a reference for the baseline algorithm that our system builds upon.


PyG optimizes performance through a feature called CPU Affinity, which allows the sampling process to utilize a pre-defined number of CPU cores via the \texttt{num\_workers} parameter. It binds worker subprocesses to CPU cores, with each worker maintaining its own data-loader and sampler state, enabling parallel batch preparation. However, since each worker operates in its own memory space, the sampler's data must be replicated across these spaces. This replication can reduce performance and limit the number of workers that can be used, especially when the sampling device has limited memory. Therefore, it is crucial to carefully assess the benefits of CPU Affinity, ensuring that the overhead introduced by parallelism is minimal compared to the gains in sampling speed. As shown in Figure~\ref{sampling_comparison} and discussed later, this is not always the case.

\subsection{Tensor Quantization}
Quantization aims to reduce data precision in a way that minimizes quantization error and preserves model accuracy. It is widely used in neural network algorithms to lower memory consumption, reduce computational complexity, and improve inference speed by reducing the precision of activations and weights \cite{ref:improving_speed_ofnn} \cite{ref:binarized} \cite{ref:deep_compression} \cite{ref:dcn_quant}. Quantization techniques have also been successfully applied to GNNs \cite{ref:gnn_quantization1}\cite{ref:gnn_quantization2}\cite{ref:gnn_quantization3}. In addition to its computational benefits, quantization can alleviate network bottlenecks in distributed systems by reducing the amount of data transmitted across the network.

\begin{figure*}[t]
  \centering
  \begin{subfigure}[t]{0.48\linewidth}
    \centering
    \includegraphics[width=\linewidth]{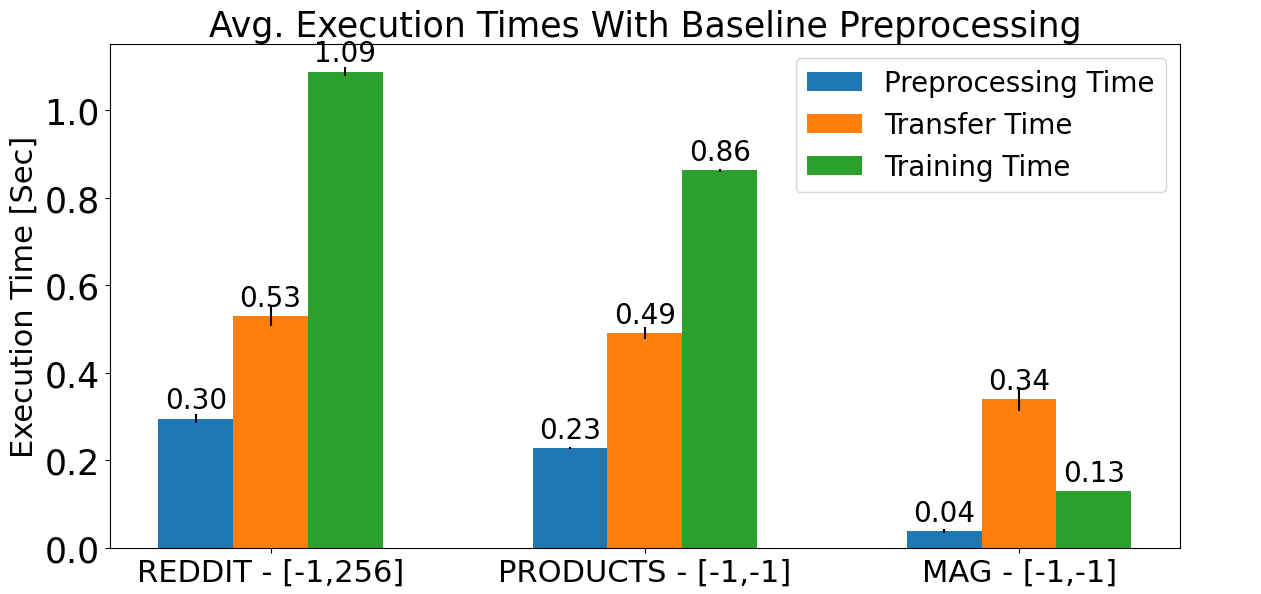}
    \caption{Mini-batch preprocessing without in-layer Neighbor Sampling}
    \label{local_comparison_left}
  \end{subfigure}
  \hfill
  \begin{subfigure}[t]{0.48\linewidth}
    \centering
    \includegraphics[width=\linewidth]{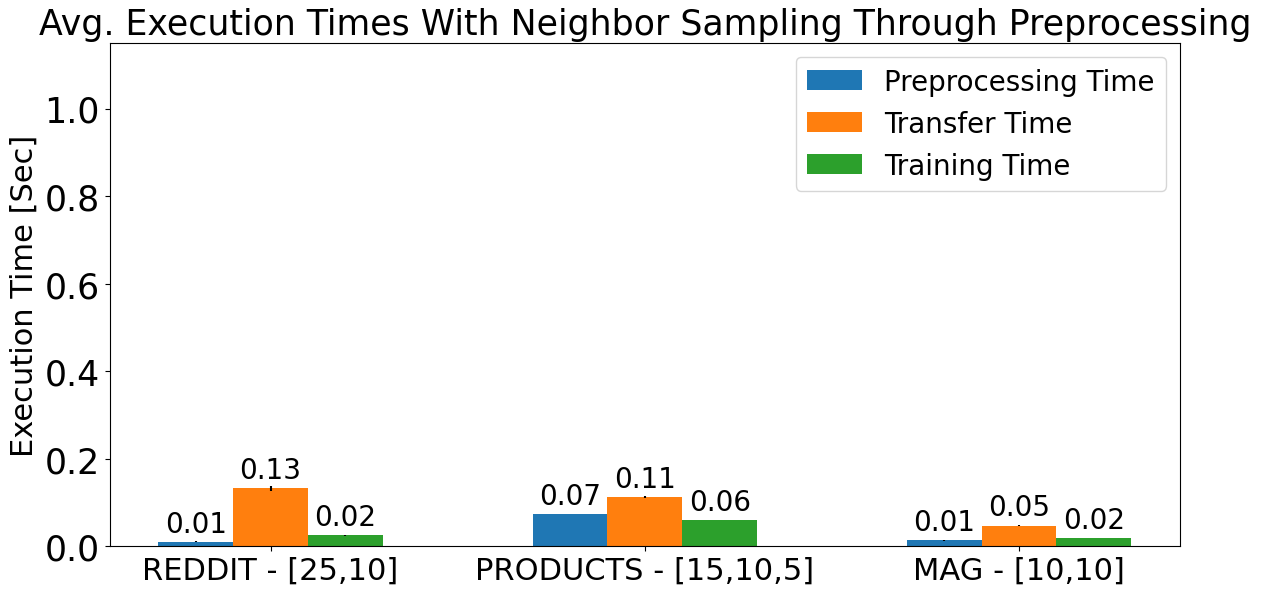}
    \caption{Preprocessing with commonly used Neighbor Sampling hyperparameters.}
    \label{local_comparison_right}
  \end{subfigure}

  \caption{Average execution times for key training phases on a baseline CPU-GPU node across all datasets.}
  \label{local_comparison}
\end{figure*}

\section{Motivation}
Our motivation for proposing LGNNIC arose from observed potential network bottlenecks in mini-batch sampling during GNN training with GraphSAGE using PyG.

GraphSAGE mini-batch training with Neighbor Sampling on a local CPU-GPU system typically involves three main phases: (1) mini-batch preprocessing, which includes Neighbor Sampling; (2) transfer of the sampled labels, nodes, edges, and their features to the GPU; and (3) GPU-based model training. To establish a baseline for later comparison with a distributed setup---comprising remote and local (i.e., computational) nodes---we measured the execution time of these three phases for each of the workloads listed in Table~\ref{tab:dataset-info}, using GraphSAGE with Neighbor Sampling as described in Subsection~\ref{subsection:Neighbor Sampling with pyg}.

Mini-batch Neighbor Sampling was configured with varying numbers of layers and well-established hyperparameters tailored to each dataset. For Reddit, we used up to 25 neighbors in the first layer and 10 in the second. For OGBN-Products, we sampled up to 15 neighbors in the first layer, 10 in the second, and 5 in the third. For OGBN-MAG, the sampling configuration included 10 neighbors in both the first and second layers. In all cases, sampling was performed on the CPU of the local node, after which the sampled mini-batches were transferred to the local GPU over the PCIe interface.

We used an AMD EPYC 7513 32-core processor and an NVIDIA A100 GPU. The PyG-based implementation of GraphSAGE mini-batch sampling \cite{ref:hamilton2018inductive} restricts the sampling operation to CPU execution. To accelerate this phase, we leveraged the CPU affinity feature and employed six worker processes via multiprocessing. For profiling, we measured the average execution times of the three main stages in each mini-batch: (1) preprocessing, (2) data transfer following sampling, and (3) GPU training.

Figure~\ref{local_comparison} compares the performance impact of using Neighbor Sampling with optimized, widely used hyperparameters (bottom) to a baseline without in-layer sampling (top) across all three phases. These hyperparameters were chosen because they are commonly used in the GNN community and yield only a minor reduction in test accuracy (as shown in Table~\ref{tab:test_accuracy_comparison} and discussed later), despite reducing the amount of data processed per batch. In the top plot, no in-layer Neighbor Sampling was applied for OGBN-Products and OGBN-MAG---i.e., while two layers were used, no neighbors were actually sampled in those layers. For the Reddit dataset, we applied no sampling in the first layer and limited sampling to 256 neighbors in the second layer to avoid an out-of-memory error due to the GPU's limited memory capacity.



Given the CPU's high processing power, preprocessing time remains relatively low, while data transfer and GPU training account for the majority of the runtime. As shown in Figure~\ref{local_comparison}, applying Neighbor Sampling significantly reduces execution time across all three training phases. This observation motivates our focus on reducing data transfer overhead by offloading Neighbor Sampling (together with Tensor Quantization) to a remote node in order to accelerate GNN training in distributed systems where communication becomes the primary bottleneck.

\section{LGNNIC System Architecture}
\begin{figure*}
  \centering
\includegraphics[width=0.7\linewidth]{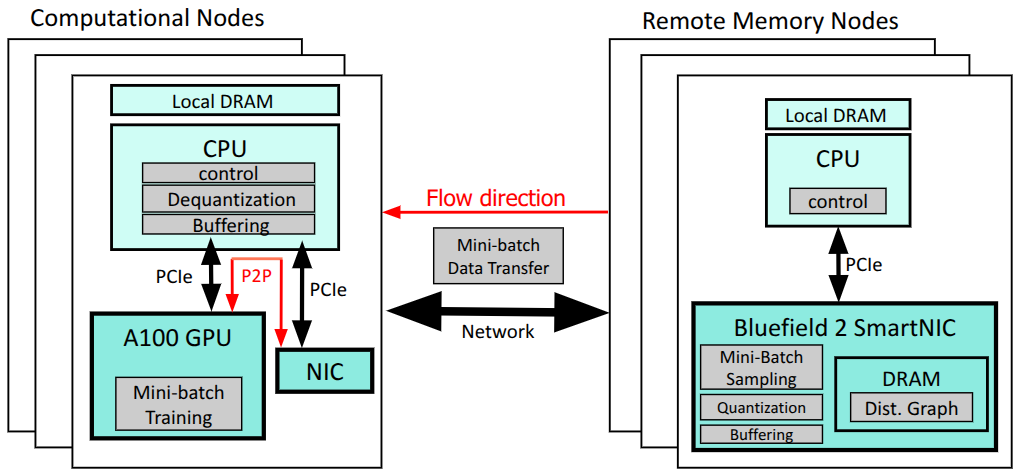}
  \vspace{-10pt}
  \caption{LGNNIC, our proposed SmartNIC-based system architecture, comprises multiple computational (local) nodes and remote memory nodes, each equipped with a SmartNIC. The graph is distributed across remote nodes, sampled and quantized on the SmartNICs before transfer.}
  \label{proposed_system}
  \vspace{-5pt}
\end{figure*}

We present our LGNNIC system architecture in Figure~\ref{proposed_system}. Our LGNNIC solution aims to accelerate the total training execution time of GNN algorithms using SmartNICs as intermediate network infrastructure. Figure~\ref{proposed_system} contains two classes of nodes: compute nodes, which perform GNN training, and remote-memory nodes, which store graph data. We assume that the two are connected by high-bandwidth Ethernet or another fabric such as InfiniBand. Each remote-memory node contains a BlueField-2 SmartNIC that accesses graph data from host memory or storage and performs mini-batch sampling and quantization before transmission. 

In the compute nodes (shown on the left in Figure~\ref{proposed_system}), GPUs such as the NVIDIA A100 or H100 reside alongside the NIC and CPU cores. Our LGNNIC architecture assumes the computational node's NIC can perform peer-to-peer data transfers directly to the host memory or the GPU's main memory. This peer-to-peer capability is available in many commercial NICs.

In remote-memory nodes (shown on the right in Figure~\ref{proposed_system}), we have BlueField-2 SmartNICs reading from host memory or storage to get the graph and performing mini-batch sampling and quantization before sending it over the network. As a result, they save transaction time by reducing the size of data being transferred. This approach mitigates network bottlenecks by distributing the full graph across multiple remote nodes and offloading mini-batch Neighbor Sampling together with Tensor Quantization closer to the remote memory nodes. 


In our architecture, to generate mini-batches that fit within the computational node’s GPU memory, the remote memory nodes collectively handle the full graph storage, with each node managing its assigned partition. These nodes execute the preprocessing phase---including both the sampling algorithm and Tensor Quantization---on their locally stored data. These preprocessed mini-batches are then transferred directly from the remote memory to the computational node's GPU, where the remainder of the training phase is conducted. While the proposed architecture is designed to scale across multiple distributed nodes, the PoC evaluated in this paper isolates a single remote-memory-node/compute-node pair to evaluate the per-node SmartNIC offloading and communication-reduction mechanisms, as detailed in Section \ref{sec:LGNNIC Evaluation}. 

\section{LGNNIC Evaluation}
\label{sec:LGNNIC Evaluation}
In this section, we begin by presenting the hardware setup and software implementations used to evaluate our proof-of-concept system, including a detailed overview of both the optimized and baseline synchronization mechanisms. We then discuss the challenges associated with offloading Neighbor Sampling from a powerful processor, such as the host CPU, to a lower-performance device like the BlueField-2 DPU. Despite these challenges, we demonstrate that offloading yields substantial benefits. Most notably, it results in a significant overall training speedup by reducing data transfers, which are a major bottleneck in high-overhead network environments. Finally, we show that Tensor Quantization results in relatively small changes in test accuracy across the evaluated datasets and hyperparameters, while yielding a marked improvement in overall training time by further decreasing the volume of transferred data.
 
\subsection{Experimental Hardware}
\label{subsec:doca}
SmartNICs combine local memory with a programmable compute engine to process data in transit across data centers. BlueField-2 \cite{ref:smartNic}, NVIDIA’s DPU-based SmartNIC, offloads infrastructure services like networking and storage. It features 8 Armv8 A72 cores, 16 GB/32 GB DDR4 memory, a PCIe Gen 4.0 intra-node connection and supports up to 200 Gb/s connectivity via Ethernet/HDR InfiniBand. BlueField is programmed using NVIDIA’s DOCA SDK \cite{ref:doca}, which provides a low-level open API for accessing its services.

\begin{figure*}[htbp]
  \centering
  \includegraphics[width=0.7\linewidth]{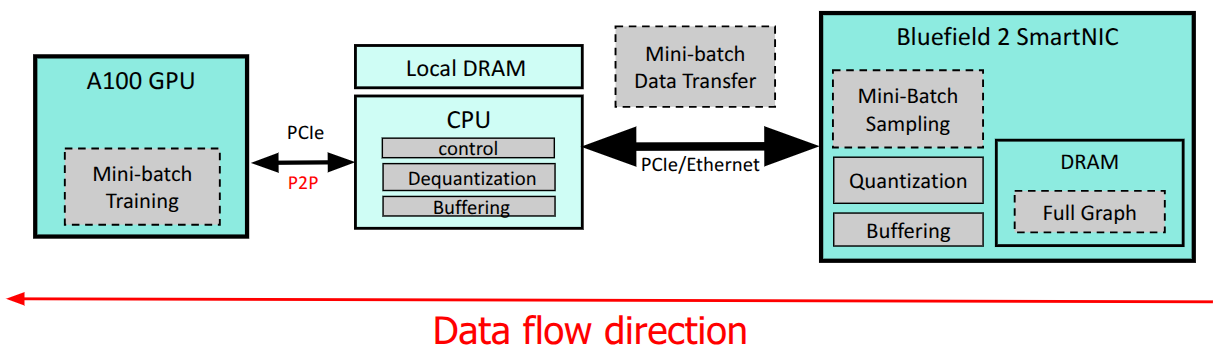}
  \caption{Our proof-of-concept system was designed to measure the execution times of key training phases using either our DOCA-DMA mechanism or the Sockets benchmark.}
  \label{PoC_system}
\end{figure*}

Our focus in this paper is to evaluate the core SmartNIC offloading mechanisms of LGNNIC using a simplified proof-of-concept system. As shown in Figure~\ref{PoC_system}, the PoC consists of one remote-memory node and one compute node and uses relatively large, widely used datasets that fit within the system's memory constraints. This configuration isolates the effects of SmartNIC-based Neighbor Sampling, Tensor Quantization, and data-transfer optimization without introducing cross-partition sampling, remote-node coordination, or multi-node routing overheads. Evaluation of these multi-node scaling effects is left to future work.

The proof-of-concept (PoC) setup used for our measurements, as shown in Figure~\ref{PoC_system}, differs from the proposed system in Figure~\ref{proposed_system}. The PoC uses a single compute and memory node to isolate and highlight the impact of the SmartNIC offloading techniques on network bottlenecks. It consists of a BlueField-2 SmartNIC, an EPYC 7513 32-core CPU, and an A100 GPU. We use the A100X, a converged card integrating a BlueField-2 DPU and an NVIDIA A100 GPU, for both components. The A100X supports two modes: one where the GPU is visible to the host system, and another where it is accessible only to the SmartNIC’s OS. Note that while Figures ~\ref{proposed_system} and ~\ref{PoC_system} depict the SmartNIC and GPU as separate discrete blocks communicating over PCIe, these diagrams represent the logical data flow rather than the physical layout of the converged A100X card.

By applying both offloading techniques, graph data is stored in the SmartNIC’s DRAM, and after in-network sampling during preprocessing, feature tensors are quantized to FP16. The sampled (and quantized) mini-batches are then transferred to the host CPU using our DOCA-DMA-based or socket-based synchronization mechanism. Finally, the mini-batches are passed to the GPU for training, including both forward and backward passes.

\subsection{Software Implementation}
To demonstrate the benefits of the architecture, we developed an optimized synchronization mechanism for executing mini-batch training with Neighbor Sampling across the PoC system. This mechanism transfers sampled mini-batches from the remote node to the computational node, managing data buffers on both ends and enabling remote-side computation during transfers. The goal is to achieve low-overhead, high-speed data transfers via DMA, while avoiding unnecessary OS-level memory copies.


To this end, our implementation leverages the low-level,
one-directional, C-based DOCA-DMA API \cite{ref:docadma}, which enables
DMA-based transfers between host and BlueField-2 DPU buffers over PCIe.
The mechanism avoids socket and network-stack overheads; however, it is
not fully zero-copy in the current implementation. Because the DOCA-DMA
buffer size is fixed and neither the buffer nor its underlying memory
pointer can be changed without reinitializing the DOCA process, received
chunks must be copied once from the host-side DOCA buffer into a larger
local aggregation buffer. Thus, each transferred chunk incurs a single
host-side memory copy before reconstruction of the corresponding tensor.

As described later in the paper, we implemented a generalized version of this API that is integrated with PyG to support a bidirectional synchronization protocol between the host (computational node) and the DPU (remote node). This protocol efficiently transfers large tensor buffers using acknowledgment (ACK) messages to ensure correctness and ordering.

\begin{figure*}[t]
\centering

\begin{minipage}[t]{0.48\textwidth}
\begin{algorithm}[H]
    \caption{BF-2 Side (Mem. node) of DOCA-DMA + PyG Training Pipeline}
    \label{alg:BlueField}
    \scriptsize
    \setlength{\baselineskip}{8pt}
    \begin{algorithmic}[1]
        \State \textbf{Initialize:} DOCA Source \& Destination Buffers
        \State \textbf{Initialize:} Python \& Python Objects
        \State \textbf{Initialize:} Dataset \& \texttt{train\_loader} Configuration
        \For{$\text{epoch} = 1$ \textbf{to} $\text{Epoch Number}$}
            \While{batch=next(\texttt{train\_loader})}
                \For {Each Batch Element Tensor} \Comment{i.e., Packet}
                    \State \textbf{Quantize:} Tensor \Comment{Optional}  
                    \State \textbf{Convert:} Batch Elements to Bytes
                    \For {Each Chunk In Element Tensor}
                        \State \textbf{Memcopy:} Data to DOCA Buffer By Order 
                        \State \textbf{Wait:} For ACK From Host
                    \EndFor
                \EndFor
            \EndWhile
            \State \textbf{Send:} Finish Message To Host
        \EndFor
    \end{algorithmic}
\end{algorithm}
\end{minipage}
\hfill
\begin{minipage}[t]{0.48\textwidth}
\begin{algorithm}[H]
    \caption{Host Side (Comp. node) of DOCA-DMA + PyG Training Pipeline}
    \label{alg:host}
    \scriptsize
    \setlength{\baselineskip}{8pt}
    \begin{algorithmic}[1]
        \State \textbf{Initialize:} DOCA Source \& Destination Buffers 
        \State \textbf{Initialize:} Python \& Python Objects
        \State \textbf{Initialize:} Model
        \State \textbf{Initialize:} Optimizer
        \For{$\text{epoch} = 1$ \textbf{to} $\text{Epoch Number}$}
            \While{Epoch\_Finished\_Message is False}
                \For {Each Batch Element Tensor} \Comment{i.e., Packet}
                    \For {Each Chunk In Element Tensor}
                        \State \textbf{Wait:} For New Chunk From BlueField-2
                        \State \textbf{Memcopy:} DOCA Buffer to Local Buffer 
                        \State \textbf{Send:} ACK To BF-2
                    \EndFor
                    \State \textbf{PyMemoryView:} Local Buffer To Tensor
                    \State \textbf{Dequantize:} Tensor \Comment{Optional} 
                \EndFor
                \If{Epoch Finished Message Was Received}
                    \State \textbf{Assign:} Epoch\_Finished\_Message = True
                    \State \textbf{Break} 
                \EndIf
                \State \textbf{Transfer:} Mini-Batch Data to GPU
                \State \textbf{Perform:} Forward Pass Through The Model
                \State \textbf{Calculate:} Loss Using Predicted and Actual Labels
                \State \textbf{Calculate:} Back-Propagate Gradients
                \State \textbf{Update:} Model Parameters Using Optimizer
            \EndWhile
        \EndFor
    \end{algorithmic}
\end{algorithm}
\end{minipage}

\end{figure*}

Integrating the C-based DOCA API with PyG’s Python-based sampling required careful coordination. To maintain buffer synchronization and performance, each sampling batch N had to be coordinated with its predecessor N-1, ensuring data from mini-batch N wouldn’t overwrite buffer contents still in use from N-1. 

Since DOCA buffers are limited to 1 MB, tensor transfers occur in chunks, requiring an ACK-based mechanism to prevent overwriting in-progress data transfers. As we further analyze in Subsection \ref{subsection: NVIDIA's BlueField-2 and DOCA}, this hardware limitation acts as a significant performance bottleneck, which directly necessitated the design of our complex chunking and synchronization protocol. The ACK$-$based synchronization mechanism sends each chunk with a header comprising the following information: 
(1) the packet number, (2) the tensor's number of bytes, (3) the number of the current chunk, and (4) the number of bytes in the current chunk (which varies between mini-batches and tensors). A packet is defined as the combination of all the chunks needed to transfer one tensor of data related to a specific sampled batch. For instance, when transferring nodes' feature tensor to the local node, a packet comprises all the chunks needed to transfer the feature tensor. Both the Reddit and OGBN-Products datasets are homogeneous, meaning they consist of only one kind of node. In addition, neither dataset includes edge features. Hence, in their case, only three packets were sent for each batch: one for the node-feature tensor, one for the batch-node labels, and one for the edges' adjacency matrix tensor, each composed of multiple chunks. However, OGBN-MAG is a heterogeneous graph, meaning it consists of more than one kind of node and edge. In addition, it does not include edge features. Hence, in OGBN-MAG's case, for each batch, twelve packets were sent with the following tensors: paper node features, paper nodes labels, author node features, institution node features, field\_of\_study features, (author, affiliated\_with, institution) adjacency matrix, (author, writes, paper) adjacency matrix, (paper, cites, paper) adjacency matrix, (paper, has\_topic, field\_of\_study) adjacency matrix, (institution, rev\_affiliated\_with, author) adjacency matrix, (paper, rev\_writes, author) adjacency matrix, and (field\_of\_study, rev\_has\_topic, paper) adjacency matrix. 

To prevent overwriting the current chunk in the local (computational) node's DOCA buffer, the remote node waits for an ACK from the local node confirming the data has been copied to a larger aggregation buffer, where all incoming tensor chunks are aggregated. Once the data is copied from its local DOCA buffer, the remote node proceeds without waiting---either handling the next chunk or sampling the next batch if the last chunk was already sent. The local node identifies new data in its DOCA buffer by detecting a change in the "current chunk number" field in the header, which is the last field to be updated by the remote node. The other header parameters are used later during training for tensor-related calculations. The synchronization flow is shown in Algorithm~\ref{alg:BlueField} for the BlueField-2 remote node and Algorithm~\ref{alg:host} for the local node.



A corresponding synchronization mechanism was also built as a benchmark with sockets to demonstrate the importance of using a low-overhead DMA-based mechanism with faster transactions and to underscore the significance of sampling, particularly when using a high-latency protocol such as Ethernet.

\subsection{Offloading Neighbor Sampling to the SmartNIC}
\subsubsection{Performance Comparison of Neighbor Sampling on Host CPU vs. BlueField-2 DPU}
\begin{figure*}[htbp]
  \centering

  \begin{subfigure}{0.32\linewidth}
    \centering
    \includegraphics[width=\linewidth]{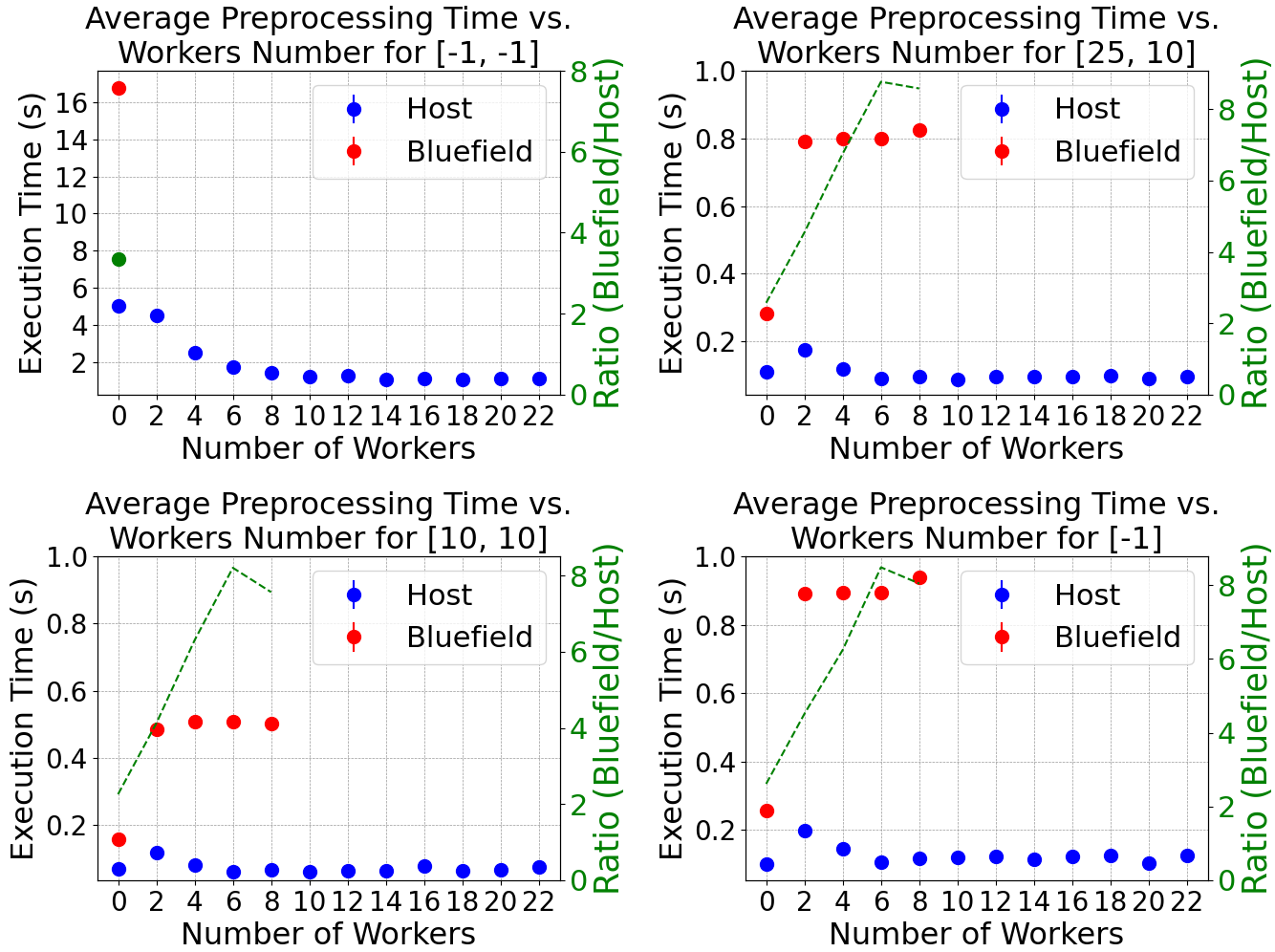}
    \caption{Reddit}
    \label{reddit_sampling_comparison}
  \end{subfigure}
  \hfill
  \begin{subfigure}{0.32\linewidth}
    \centering
    \includegraphics[width=\linewidth]{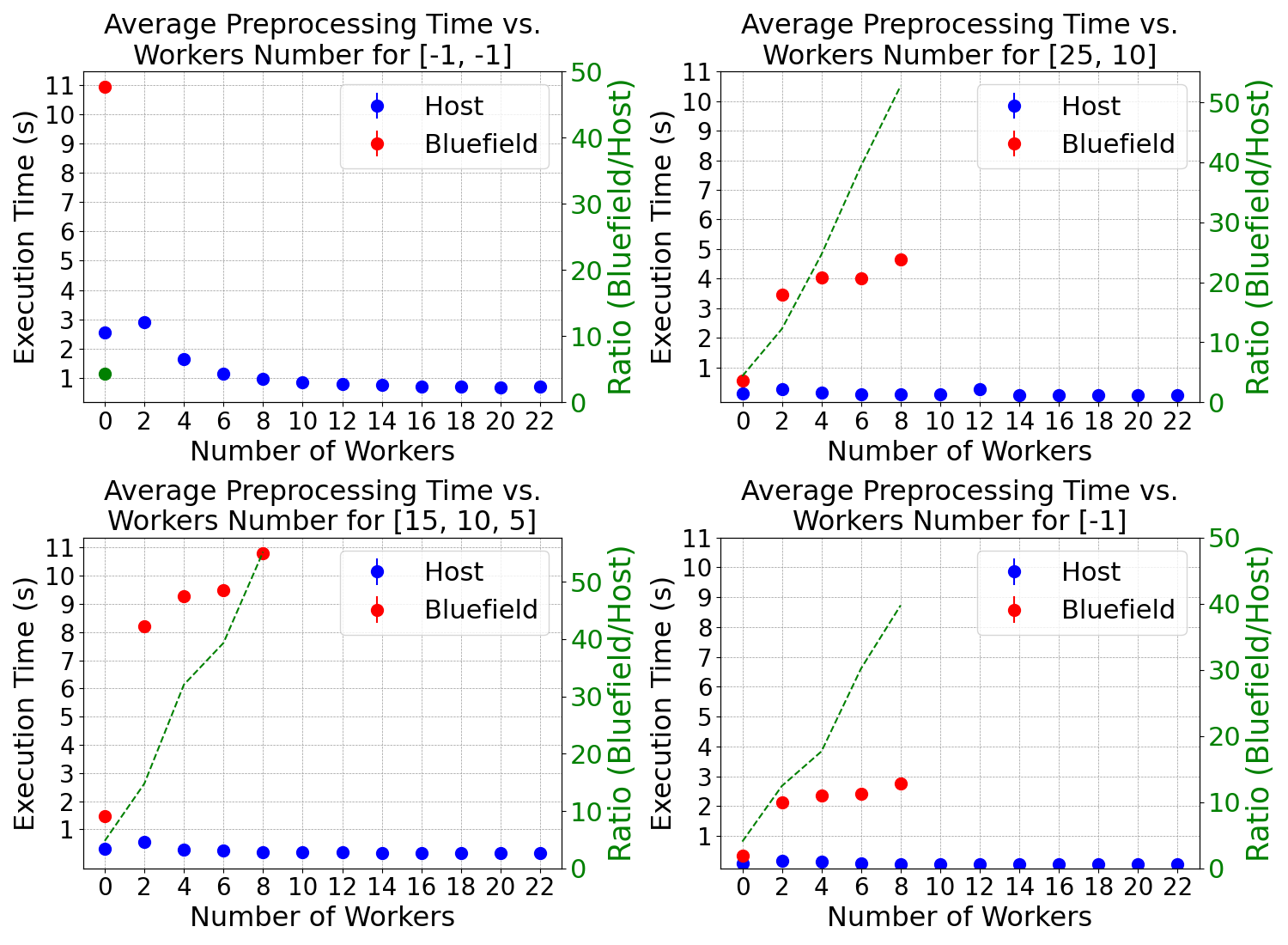}
    \caption{OGBN-Products}
    \label{products_sampling_comparison}
  \end{subfigure}
  \hfill
  \begin{subfigure}{0.32\linewidth}
    \centering
    \includegraphics[width=\linewidth]{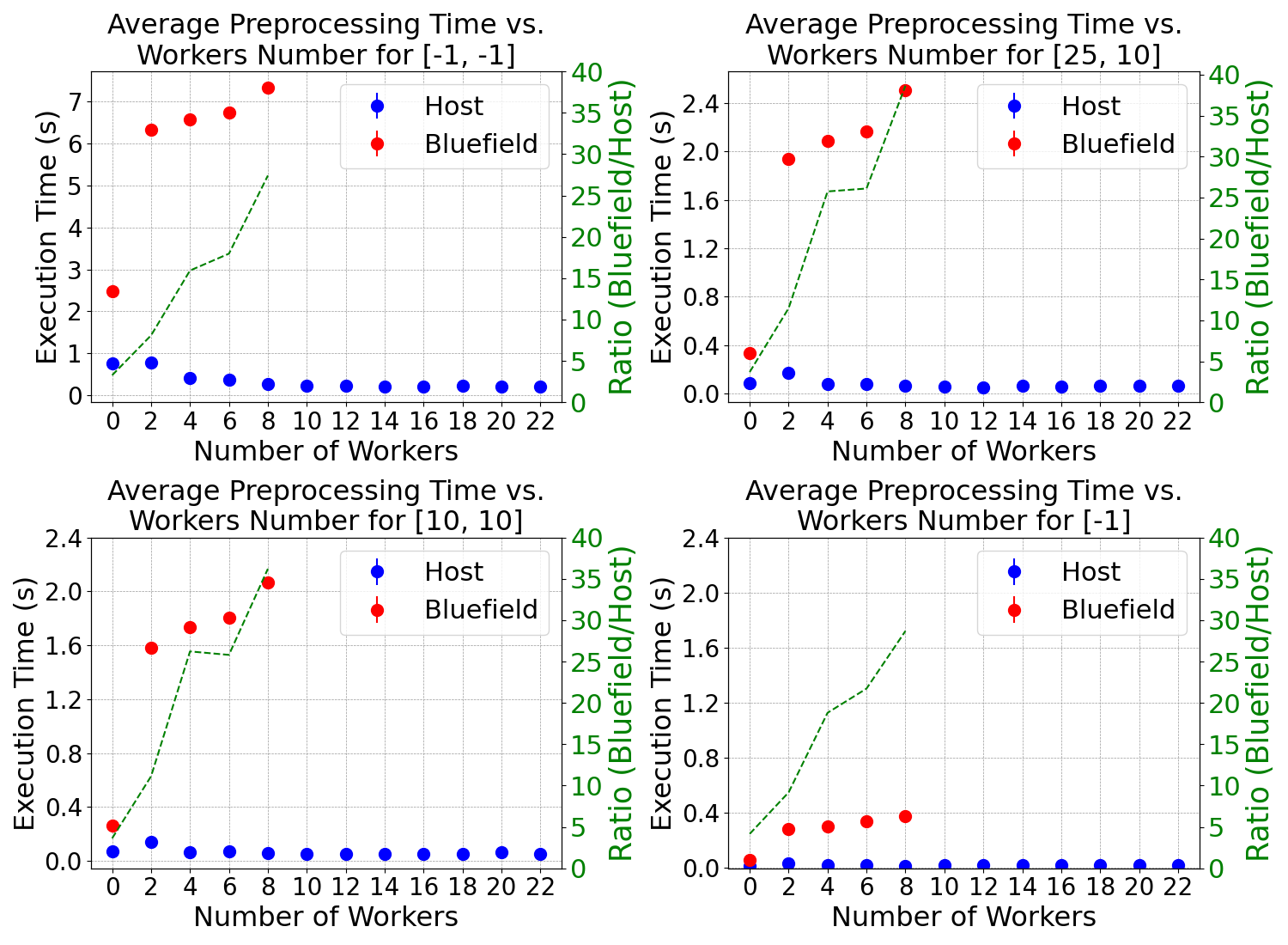}
    \caption{OGBN-MAG}
    \label{mag_sampling_comparison}
  \end{subfigure}

  \caption{Impact of in-layer Neighbor Sampling on preprocessing performance, comparing execution on the host CPU versus the BlueField-2 SmartNIC across the three datasets under various sampling hyperparameters.}
  \label{sampling_comparison}
\end{figure*}

To evaluate the computational disparities between our BlueField-2 SmartNIC and the powerful local node EPYC CPU---two devices whose performance difference reflects realistic system architectures---we compared the execution time of preprocessing two layers in the computational trees with and without in-layer Neighbor Sampling (and other configurations) on the host CPU versus the remote BlueField-2 node. Quantization was not applied in this part of the evaluation. The measurements were taken as a function of the number of workers for each dataset: Reddit, OGBN-Products, and OGBN-MAG, as illustrated in Figure~\ref{sampling_comparison}.

The mini-batch preprocessing phase (top-left subfigures in Figures \ref{reddit_sampling_comparison}, \ref{products_sampling_comparison}, \ref{mag_sampling_comparison}) entails constructing two-layer computational trees for each node in the mini-batch without applying sampling at these layers (i.e., no in-layer sampling). Additionally, for each dataset, three mini-batch preprocessing steps including extensive sampling with varying hyperparameters were evaluated. For all datasets, not only were the commonly used hyperparameters employed, but additional ones were tested to explore the effects of Neighbor Sampling with varying numbers of nodes and layers. Each hyperparameter is enclosed in brackets, [], where each number represents the maximum number of nodes to sample at a given layer. Reading from left to right, the first number corresponds to the sampling size for the first neighbor hop from the root in its computational tree, continuing sequentially for subsequent layers. If \textit{L} numbers are specified, this indicates that the computational tree consists of \textit{L} layers (hops). When a value of -1 is used, it indicates that no sampling was performed at that layer, meaning all neighbors in that layer were retained.

For the Reddit dataset, the following sampling hyperparameters were selected and are shown in Figure~\ref{reddit_sampling_comparison}: [-1, -1] (no in-layer sampling, top left), [25, 10] (commonly used hyperparameters, top right), [10, 10] (more extensive sampling, bottom left), and [-1] (a single layer with no in-layer sampling, bottom right). For the OGBN-Products dataset, the sampling hyperparameters were as follows and are presented in Figure~\ref{products_sampling_comparison}: [-1, -1] (no in-layer sampling, top left), [25, 10] (more extensive sampling, top right), [15, 10, 5] (commonly used hyperparameters, bottom left), and [-1] (a single layer with no in-layer sampling, bottom right). For the OGBN-MAG dataset, the selected sampling hyperparameters, shown in Figure~\ref{mag_sampling_comparison}, were: [-1, -1] (no in-layer sampling, top left), [25, 10] (more extensive sampling, top right), [10, 10] (commonly used hyperparameters, bottom left), and [-1] (a single layer with no in-layer sampling, bottom right).

Specifically, on the SmartNIC, configurations combining CPU Affinity (\(\texttt{num\_workers} > 0\)) with unconstrained two-hop Neighbor Sampling (\(\texttt{num\_neighbors}=[-1, -1]\), i.e., all neighbors were retained at both hops) exhausted the available DRAM capacity across all evaluated datasets. Memory monitoring showed a rapid increase in memory consumption until the device's available memory was depleted. Because each worker executes in a separate process and replicates the data loader and sampler state, multi-worker execution substantially increases the aggregate memory footprint on the BlueField-2 (Figures~\ref{reddit_sampling_comparison}, \ref{products_sampling_comparison}, and \ref{mag_sampling_comparison}). Notably, increasing the number of unconstrained neighborhood-expansion layers exhausted the available SmartNIC memory even with \(\texttt{num\_workers}=0\). The corresponding runs did not complete successfully and were therefore excluded from the performance figures.

\begin{figure*}[htbp]
  \centering

  \begin{subfigure}{0.32\linewidth}
    \centering
    \includegraphics[width=\linewidth]{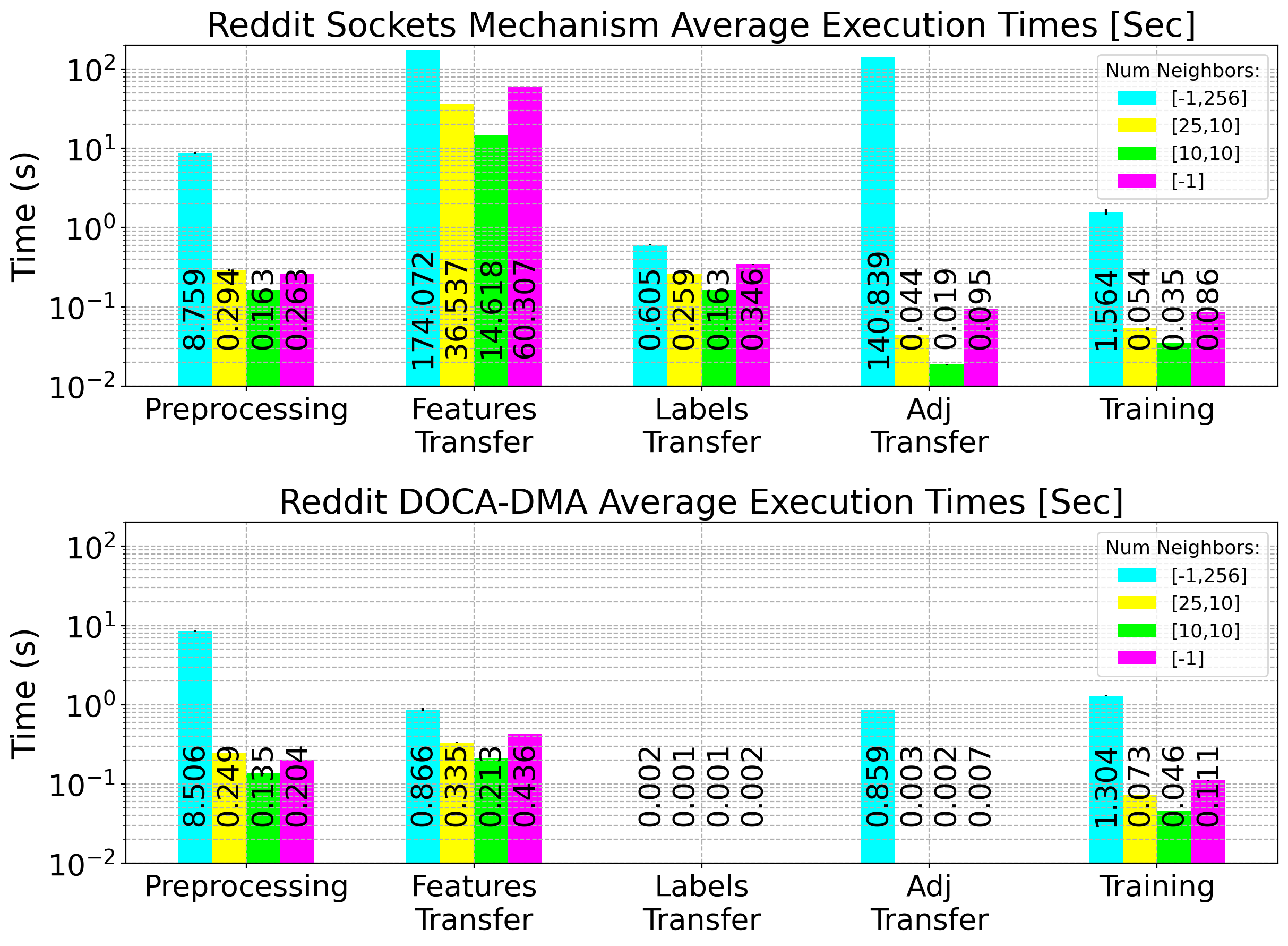}
    \caption{Reddit}
    \label{reddit_sockets_vs_doca}
  \end{subfigure}
  \hfill
  \begin{subfigure}{0.32\linewidth}
    \centering
    \includegraphics[width=\linewidth]{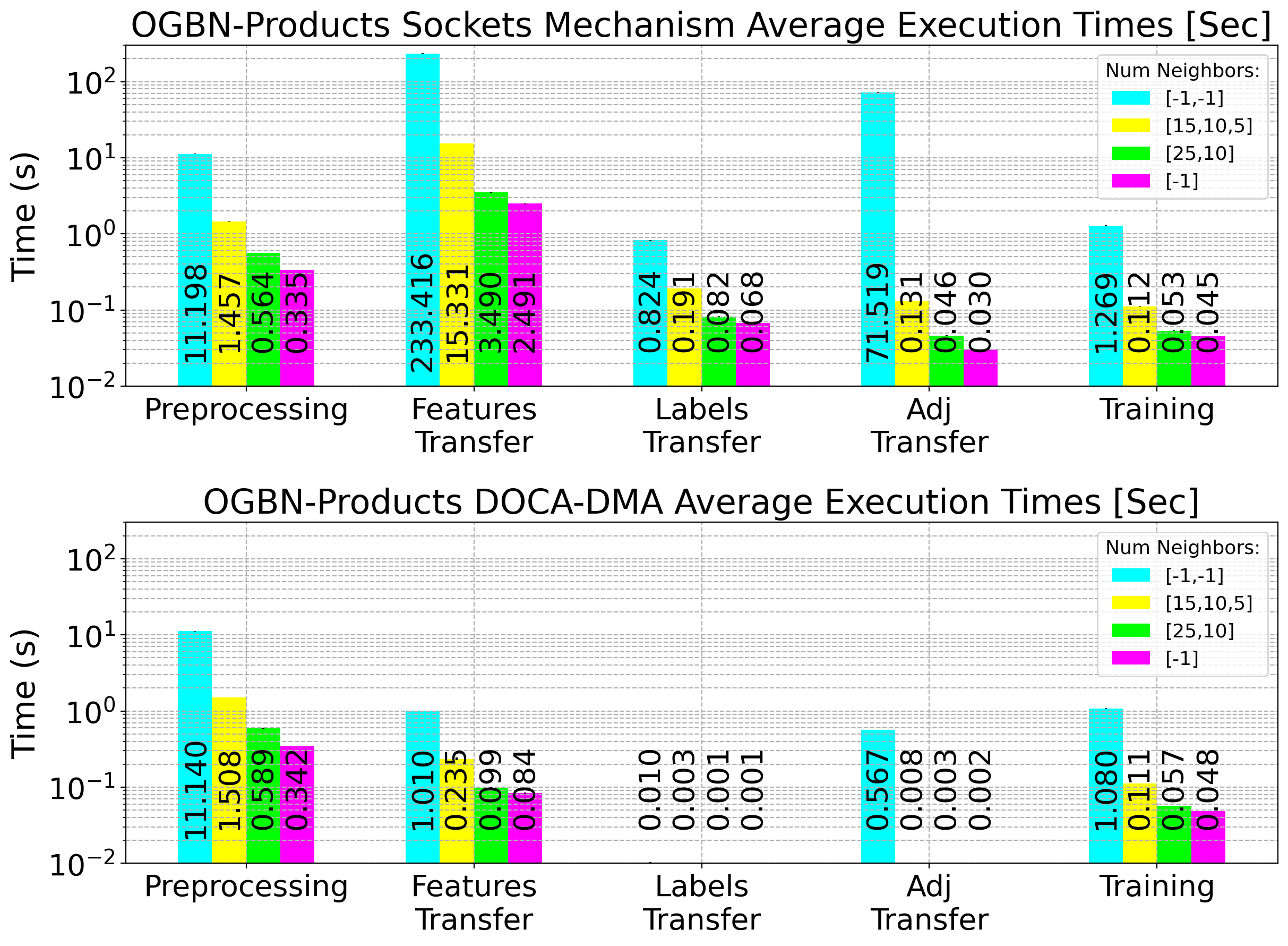}
    \caption{OGBN-Products}
    \label{ogbn_products_sockets_vs_doca}
  \end{subfigure}
  \hfill
  \begin{subfigure}{0.32\linewidth}
    \centering
    \includegraphics[width=\linewidth]{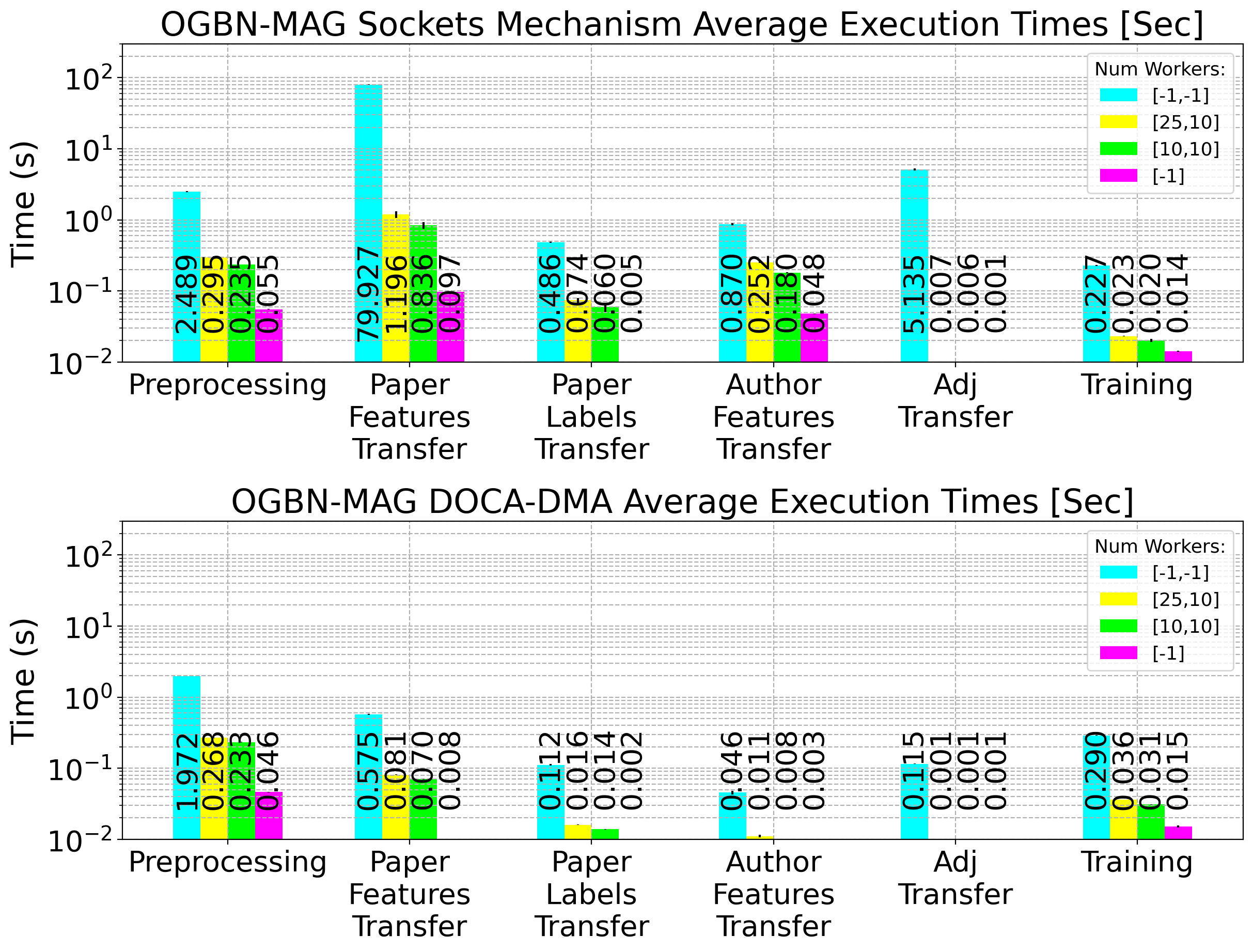}
    \caption{OGBN-MAG}
    \label{ogbn_mag_sockets_vs_doca}
  \end{subfigure}

  \caption{Average execution time comparison across all datasets. Key mini-batch training phases are analyzed using various sampling hyperparameters with our mechanisms to demonstrate the impact of Neighbor Sampling.}
  \label{sockets_vs_doca}
  \vspace{-10pt}
\end{figure*}

Increasing the number of workers did not improve preprocessing
performance on the SmartNIC and degraded it in several configurations.
Even under the most favorable configurations with extensive sampling,
preprocessing on the BlueField-2 remained slower than on the
EPYC~7513 CPU because of its lower computational capability and more
limited memory resources. With \(\texttt{num\_workers}=0\),
BlueField-2 Neighbor Sampling was approximately
\(2\text{--}10\times\) slower than sampling on the EPYC~7513.

Nevertheless, as evidenced by Figures \ref{reddit_sampling_comparison}, \ref{products_sampling_comparison}, and \ref{mag_sampling_comparison}, extensive sampling significantly reduces preprocessing time for both the CPU and SmartNIC compared to the no-in-layer scenario, especially for the SmartNIC when CPU Affinity is disabled, as it reduces memory access demands. Therefore, as demonstrated in Figures \ref{reddit_sampling_comparison}, \ref{products_sampling_comparison}, and \ref{mag_sampling_comparison}, since preprocessing on the SmartNIC without CPU Affinity provided superior performance, we disabled this feature (\( \texttt{num\_workers} = 0 \)).

As shown in Figure~\ref{sampling_comparison}, the SmartNIC demonstrates a significant preprocessing time reduction when utilizing in-layer Neighbor Sampling (comparing the top-left subfigures to the commonly used hyperparameters in each dataset with \( \texttt{num\_workers} = 0 \)). For example, with Reddit (Figure~\ref{reddit_sampling_comparison}), BlueField's preprocessing time dropped from $\sim17$ seconds to $\sim0.3$ seconds. Other datasets show similar results.



\subsubsection{Overall Training Speedup through Offloaded Neighbor Sampling}
While Figure~\ref{sampling_comparison} highlights the difficulty of sampling on Blue\-Field-2, Figure~\ref{sockets_vs_doca} shows that network bottlenecks can dominate execution time, making data transfer time (from SmartNIC to host) a key factor. As shown later, performing sampling on the SmartNIC can significantly reduce transfer time---even when sampling itself is slow---underscoring the SmartNIC’s value in accelerating overall training.

As illustrated in Figure~\ref{sockets_vs_doca}, performing remote preprocessing with sampling not only decreases sampling time but also significantly reduces the transfer time for the preprocessed batches. To showcase the benefits of the architecture, we implemented a DOCA-DMA-based synchronization mechanism to transfer the preprocessed batches from the remote node to the local node. This mechanism enables low-overhead, high-speed transactions via DMA, eliminating unnecessary OS memory copies and allowing concurrent preprocessing of the next batch while the current one is in transit. For comparison, we also developed a socket-based synchronization mechanism to expose the network bottleneck that arises when using Ethernet instead of PCIe.

Figure~\ref{sockets_vs_doca} contrasts the performance of the two synchronization mechanisms---DOCA-DMA (lower subfigure in Figures \ref{reddit_sockets_vs_doca}, \ref{ogbn_products_sockets_vs_doca}, and \ref{ogbn_mag_sockets_vs_doca}) and Sockets (upper subfigure in Figures \ref{reddit_sockets_vs_doca}, \ref{ogbn_products_sockets_vs_doca}, and \ref{ogbn_mag_sockets_vs_doca})---on our proof-of-concept system. The subfigures present execution time measurements for various training phases across all datasets using different sampling hyperparameters mentioned in each subfigure's legend. Notably, for Reddit, we selected the [-1, 256] configuration (which represents no sampling in the first layer and sampling up to 256 neighbors in the second layer) instead of [-1, -1] due to the memory constraints of the A100 GPU (80 GB), as removing the neighbor limit would exceed GPU capacity. The [-1, 256] setting provided the minimal sampling necessary to stay within the GPU’s memory limits.

For each dataset, we measured three training phases: preprocessing on the SmartNIC, transfer of feature, label, and adjacency matrix tensors (at least one each) from the SmartNIC to the local node, and training on the local node. Each transaction consists of a single packet divided into multiple chunks (each chunk includes the relevant tensor bytes and four header fields). This segmentation is due to the 1 MB buffer limit of the DOCA-DMA library, and for consistency, we set the Sockets mechanism’s buffer size to 1 MB as well, which limits the performance of the Sockets mechanism. Using our synchronization mechanisms, tensor chunks are transferred from the SmartNIC to the local CPU for GPU training. Since each dataset consists of multiple tensors representing different features, labels, and adjacency matrices, only non-negligible data transfer times are shown in Figure~\ref{sockets_vs_doca} due to space constraints. However, every individual tensor transfer time was measured and included in the calculation of both the total training time speedup and total transfer time speedup, as shown in Table~\ref{tab:acceleration_results}. 

As illustrated in Figure~\ref{sockets_vs_doca}, there is a significant reduction in the execution times of the three phases across all datasets. For example, for the Reddit dataset with the Sockets mechanism, the average feature tensor transfer time decreases from 174.072 seconds (for the [-1, 256] hyperparameter) to 36.537 seconds (for the commonly used hyperparameter [25, 10]). Other datasets show similar results.

Figure~\ref{sockets_vs_doca} shows that as sampling becomes more extensive, transfer times drop significantly due to reduced data volume. However, while sampling lowers preprocessing time, it is not negligible and can become a bottleneck as transfer costs diminish. As shown in Table~\ref{tab:dataset-info}, the Reddit dataset has far more edges---about 1.85x more than OGBN-Products and 5.5x more than OGBN-MAG---despite having far fewer nodes (roughly 10.5x fewer than OGBN-Products and 8.3x fewer than OGBN-MAG). Thus, sampling batches of 1024 nodes yields longer transfer times for Reddit, as each computational tree includes more nodes on average. This results in extended transfer times for Reddit’s feature and adjacency tensors under identical sampling hyperparameters for both DOCA-DMA and Socket mechanisms. Additionally, the commonly used OGBN-Products hyperparameter ([15, 10, 5]) employs three sampling layers, causing higher preprocessing time than other extensive sampling settings---both within OGBN-Products and compared to other datasets.


To gain deeper insights into the results, we calculated the speedup of extensive sampling---using various hyperparameters, including those commonly used to achieve optimal accuracy---compared to the baseline without in-layer sampling (i.e., [-1, -1] or [-1, 256]) for each dataset. The speedup in transaction time and total training time for both DOCA-DMA and Sockets is substantial across all datasets and hyperparameters, as shown in Table~\ref{tab:acceleration_results}. For each dataset, the transaction speedup was calculated by dividing the average total transfer time for batches with minimal sampling ([-1, 256] or [-1, -1], depending on the dataset) by the average total transfer time for batches with extensive sampling ([25, 10], [10, 10]/[15, 10, 5], [-1]). The results for each dataset are illustrated in Table~\ref{tab:acceleration_results}.

Total training time speedup was calculated in a similar manner, taking into account preprocessing and training phases. The results, summarized in Table~\ref{tab:acceleration_results} alongside Figure~\ref{sockets_vs_doca}, highlight that graph preprocessing through sampling can mitigate network bottlenecks and significantly reduce data transfer times across all datasets. Since not all hyperparameter combinations are applicable to all datasets, the corresponding cells are left empty and marked with a ‘-’ sign.

\begin{table*}[htbp]
\centering
\scriptsize
\resizebox{\textwidth}{!}{
\begin{tabular}{|
  >{\columncolor{gray!20}}c|c|c|c|c|c|c|
  >{\columncolor{gray!20}}c|c|c|c|c|c|c|
  }
\hline
\rowcolor{gray!30}
& \multicolumn{6}{c|}{\textbf{Sockets}} & \multicolumn{1}{c|}{} & \multicolumn{6}{c|}{\textbf{DOCA-DMA}} \\
\hline
\rowcolor{gray!20}
\textbf{Num Neigh.} &
\multicolumn{3}{c|}{\textbf{Trans. Speedup}} & \multicolumn{3}{c|}{\textbf{Total Speedup}} &
& \multicolumn{3}{c|}{\textbf{Trans. Speedup}} & \multicolumn{3}{c|}{\textbf{Total Speedup}} \\
\cline{2-7}\cline{9-14}
\rowcolor{gray!20}
& Reddit & Products & MAG & Reddit & Products & MAG & & Reddit & Products & MAG & Reddit & Products & MAG \\
\hline

[25, 10] &
\cellcolor{yellow}8.56  & 84.49  & 52.64 & \cellcolor{yellow}8.76 & 75.12 & 45.51 &
& \cellcolor{yellow}5.10  & 15.36  & 7.44   & \cellcolor{yellow}17.46 & 18.43  & 7.45 \\
\hline

[10, 10] &
21.32  & -- & \cellcolor{yellow}73.57  & 21.73 & -- & \cellcolor{yellow}62.39 &
& 7.99 & -- & \cellcolor{yellow}8.73 & 29.03 & -- & \cellcolor{yellow}8.61 \\
\hline

[15, 10, 5] &
-- & \cellcolor{yellow}19.53  & --  & -- & \cellcolor{yellow}18.48  & -- &
& -- & \cellcolor{yellow}6.45 & -- & -- & \cellcolor{yellow}7.40 & -- \\
\hline

[-1] &
5.19 & 118.12 & 472.08 & 5.33 & 107.17 & 353.57 &
& 3.88 & 18.18  & 47.65  & 15.20 & 28.89  & 39.40 \\
\hline

\end{tabular}
} 
\caption{Total transaction and training speedups under different sampling hyperparameters and synchronization mechanisms. Highlighted values correspond to the most common configuration per dataset.}
\label{tab:acceleration_results}
\end{table*}

Figure~\ref{sockets_vs_doca} shows that DOCA-DMA consistently outperforms the socket-based mechanism across all datasets, primarily due to its use of PCIe instead of Ethernet. Table~\ref{tab:acceleration_results} further highlights this contrast: while DOCA-DMA is faster overall, the relative transaction speedup is greater for the socket-based method, since Ethernet's higher overhead makes reductions in data transfer time more impactful. For Reddit, the reported speedup is smaller because it is measured relative to the minimal sampling setting ([-1, 256]) rather than a potential full no in-layer sampling configuration ([-1, -1]), as discussed earlier.


Additionally, for the Sockets mechanism, when data transfers are the dominant bottleneck, as seen in the Reddit Sockets measurements, the total time acceleration closely aligns with the transaction speedup. In contrast, when preprocessing and training phases significantly contribute to the overall training time, the influence of transaction speedup on total training time acceleration diminishes, as seen with Reddit’s DOCA-DMA results. In such cases, optimizing the preprocessing phase becomes crucial. As shown in Figure~\ref{sampling_comparison}, employing a more powerful CPU can substantially lower preprocessing times, highlighting the need for further exploration of hardware-based optimizations.

Regardless of the underlying reason, reducing the preprocessing or training phases through sampling, or minimizing transaction times due to sampling, ultimately results in a shorter total execution time. In summary, by synthesizing the findings from Figure~\ref{sockets_vs_doca} and Table~\ref{tab:acceleration_results}, we demonstrate that remote sampling effectively reduces transaction times, alleviates network bottlenecks, and thereby enhances overall system performance.

\subsection{Tensor Quantization on the SmartNIC}
As shown in Figure~\ref{sockets_vs_doca}, transaction time increases with the amount of data transferred over the network. In the previous section, we demonstrated that offloading Neighbor Sampling to the SmartNIC significantly accelerates data transfers by reducing the data volume. Similarly, quantization can further reduce data size and is a lightweight task for the SmartNIC, whether implemented in software or hardware. As illustrated in Figure~\ref{sockets_vs_doca}, the feature tensor consistently emerges as a primary bottleneck, accounting for a substantial portion of the transfer time across all datasets, sampling hyperparameters, and synchronization mechanisms. To address this, we utilize the SmartNIC to quantize the feature tensor from FP32 to FP16 in software, thereby reducing its size, lowering transfer time, and shortening total training time (see Table~\ref{tab:acceleration_results_quantization}).

\subsubsection{Impact of Feature Tensor Quantization on Test Accuracy}
While quantization reduces tensor size, it can also introduce error that may degrade test accuracy. Since FP32 is the default precision in many PyG datasets and models, FP16 offers a practical trade-off by minimizing quantization error while enabling considerable speedups. As shown in Table~\ref{tab:test_accuracy_comparison}, this precision reduction leads to only minor accuracy degradation, if any, while significantly accelerating data movement and improving overall training time as illustrated in Table~\ref{tab:acceleration_results_quantization}.

\begin{table}[t]
\centering
\scriptsize
\resizebox{\columnwidth}{!}{%
\begin{tabular}{|>{\columncolor{gray!20}}c|c|c|c|c|c|c|}
\hline
\rowcolor{gray!30}
\textbf{\shortstack{Num\\Neighbors}} 
& \multicolumn{2}{c|}{\textbf{Reddit}} 
& \multicolumn{2}{c|}{\textbf{\shortstack{OGBN-\\Products}}} 
& \multicolumn{2}{c|}{\textbf{\shortstack{OGBN-\\MAG}}} \\
\cline{2-7}
\rowcolor{gray!30}
& FP32 & FP16 & FP32 & FP16 & FP32 & FP16 \\
\hline
{[-1, -1]}     & --     & --     & 0.777 & 0.774 & 0.445 & 0.444 \\
{[-1, 256]}    & 0.951 & 0.951 & --     & --     & --     & -- \\
{[15, 10, 5]}   & --     & --     & \cellcolor{yellow}0.789 & \cellcolor{yellow}0.790 & --     & -- \\
{[25, 10]}     & \cellcolor{yellow}0.952 & \cellcolor{yellow}0.952 & 0.771 & 0.776 & 0.444 & 0.447 \\
{[10, 10]}     & 0.952 & 0.952 & --     & --     & \cellcolor{yellow}0.453 & \cellcolor{yellow}0.451 \\
{[-1]}        & 0.943 & 0.938 & 0.685 & 0.673 & 0.394 & 0.381 \\
\hline
\end{tabular}
}
\caption{Average test accuracy results across all datasets and sampling hyperparameters with the feature tensor quantized from FP32 to FP16. Highlighted values represent common configurations per dataset.}
\label{tab:test_accuracy_comparison}
\end{table}

In Table~\ref{tab:test_accuracy_comparison}, we report the average test accuracy for each dataset and each Num Neighbors hyperparameter (with commonly used values highlighted in yellow), comparing FP32 and FP16 precisions. Since not all hyperparameter combinations are applicable to all datasets, the corresponding cells are left empty and marked with a ‘-’ sign. As shown, the accuracy degradation from quantization is relatively small across the evaluated configurations and well justified by the resulting performance gains as illustrated in Table~\ref{tab:acceleration_results_quantization}. Each training run was repeated four times. During each run, test accuracy was evaluated using the model configuration that achieved the best validation accuracy, computed over multiple epochs until convergence. The final reported test accuracy represents the average across all four runs.      

\subsubsection{Overall Training Speedups through Offloading Tensor Quantization}
In Table~\ref{tab:acceleration_results_quantization}, we compare FP32 and FP16 precision when quantizing the feature tensor to highlight the impact of quantization on total transfer time and, consequently, total training (execution) time. Average total transaction time, average total execution time, and the resulting speedups were calculated across all mechanisms (DOCA-DMA and Sockets), datasets, and sampling hyperparameters (with common hyperparameters highlighted in yellow). Regarding Table~\ref{tab:acceleration_results_quantization}, total transaction time is defined as the sum of all data transfers from the SmartNIC to the host across all mini-batches, averaged over multiple runs. Total execution time is the sum of the three main phases: (1) preprocessing, (2) all data transfers from the SmartNIC to the host, and (3) GPU-based training, also averaged across multiple runs. Speedup values for transfer and training times were computed by dividing FP32 results by their corresponding FP16 results and are shown in Table~\ref{tab:acceleration_results_quantization}.

The different hyperparameters used in the measurements affect the size of the feature tensor transferred across the network and thus allow for evaluation under a range of data transfer conditions. Hence, since speedups are computed with the sampling hyperparameter held constant (and only the precision varied), the relative gains are solely attributed to the change in precision. This differs from Table~\ref{tab:acceleration_results}, where speedups were computed for each FP32 configuration relative to the baseline hyperparameter with minimal sampling (i.e., [-1, 256] or [-1, -1], depending on the dataset).

\begin{table*}[h!]
\centering
\resizebox{\textwidth}{!}{%
\begin{tabular}{c c c}

\begin{tabular}{|>{\columncolor{gray!20}}c|>{\columncolor{gray!20}}c|c|c|}
\hline
\rowcolor{gray!20}
\cellcolor{white}\textbf{REDDIT}& \makecell{\textbf{Num}\\\textbf{Neighbors}}
& \makecell{\textbf{Trans.}\\\textbf{Speedup}}
& \makecell{\textbf{Exec.}\\\textbf{Speedup}} \\
\hline
& \([-1, 256]\) & 1.64 & 1.62 \\
\textbf{Sockets} & \([25, 10]\) & \cellcolor{yellow}3.64 & \cellcolor{yellow}3.56 \\
& \([10, 10]\) & 3.60 & 3.48 \\
& \([-1]\) & 3.65 & 3.59 \\
\hline
& \([-1, 256]\) & 1.24 & 1.03 \\
\makecell{\textbf{DOCA}\\\textbf{-DMA}} & \([25, 10]\) & \cellcolor{yellow}1.36 & \cellcolor{yellow}1.20 \\
& \([10, 10]\) & 1.24 & 1.15 \\
& \([-1]\) & 1.37 & 1.24 \\
\hline
\end{tabular}

& 

\begin{tabular}{|>{\columncolor{gray!20}}c|>{\columncolor{gray!20}}c|c|c|}
\hline
\rowcolor{gray!20}
\cellcolor{white}\makecell{\textbf{OGBN-}\\\textbf{PRODUCTS}}& \makecell{\textbf{Num}\\\textbf{Neighbors}}
& \makecell{\textbf{Trans.}\\\textbf{Speedup}}
& \makecell{\textbf{Exec.}\\\textbf{Speedup}} \\
\hline
& \([-1, -1]\) & 2.25 & 2.16 \\
\textbf{Sockets} & \([15, 10, 5]\) & \cellcolor{yellow}3.39 & \cellcolor{yellow}2.84 \\
& \([25, 10]\) & 3.63 & 2.71 \\
& \([-1]\) & 4.53 & 3.31 \\
\hline
& \([-1, -1]\) & 1.24 & 1.02 \\
\makecell{\textbf{DOCA}\\\textbf{-DMA}} & \([15, 10, 5]\) & \cellcolor{yellow}1.41 & \cellcolor{yellow}1.04 \\
& \([25, 10]\) & 1.18 & 1.03 \\
& \([-1]\) & 2.02 & 1.11 \\
\hline
\end{tabular}
& 

\begin{tabular}{|>{\columncolor{gray!20}}c|>{\columncolor{gray!20}}c|c|c|}
\hline
\rowcolor{gray!20}
\cellcolor{white}\makecell{\textbf{OGBN-}\\\textbf{MAG}}& \makecell{\textbf{Num}\\\textbf{Neighbors}}
& \makecell{\textbf{Trans.}\\\textbf{Speedup}}
& \makecell{\textbf{Exec.}\\\textbf{Speedup}} \\
\hline
& \([-1, -1]\) & 3.03 & 2.95 \\
\textbf{Sockets} & \([25, 10]\) & 2.16 & 2.01 \\
& \([10, 10]\) & \cellcolor{yellow}1.96 & \cellcolor{yellow}1.83 \\
& \([-1]\) & 1.20 & 1.20 \\
\hline
& \([-1, -1]\) & 1.33 & 1.13 \\
\makecell{\textbf{DOCA}\\\textbf{-DMA}} & \([25, 10]\) & 1.93 & 1.25 \\
& \([10, 10]\) & \cellcolor{yellow}1.96 & \cellcolor{yellow}1.31 \\
& \([-1]\) & 1.12 & 1.05 \\
\hline
\end{tabular}

\end{tabular}%
}

\caption{Speedups achieved using FP32 vs. FP16 precision: total transaction and execution time improvements across all mechanisms, datasets, and sampling hyperparameters. For more data, see Appendix~\ref{subsection: appendix: Overall Training Speedup through Offloading Tensor Quantization}.}

\label{tab:acceleration_results_quantization}
\end{table*}

We evaluate performance under both DOCA-DMA and Sockets synchronization mechanisms to assess how different infrastructures and communication protocols impact end-to-end training. As shown in Table~\ref{tab:acceleration_results_quantization}, quantization consistently delivers significant speedups in both total transfer time and total training time across all datasets, sampling configurations, and mechanisms. Notably, in settings where transaction time dominates overall training time---such as with the high-overhead Sockets mechanism---the execution speedup closely tracks the transfer speedup. This effect is particularly pronounced in the Reddit dataset, which has the highest edge-to-node ratio among those evaluated. Detailed results are provided in Appendix \ref{subsection: appendix: Overall Training Speedup through Offloading Tensor Quantization} (Table~\ref{tab:acceleration_results_quantization_appendix}). 

The high edge-to-node ratio also explains why Reddit exhibits the highest total training time among the evaluated datasets with Sockets, while preprocessing time remains moderate. In contrast, when transaction time constitutes a smaller fraction of total training time---as is typically the case with the low-overhead DOCA-DMA mechanism, where preprocessing is also a significant bottleneck---the execution and transfer speedups diverge. Since DOCA-DMA reduces communication overhead substantially, the impact of quantization is less pronounced than with Sockets, where communication overhead remains the primary bottleneck.

In summary, this section demonstrates that quantizing the feature tensor to FP16 is an effective strategy for accelerating training while incurring minimal test accuracy degradation. Even though our quantization was deliberately simple, the SmartNIC is a natural platform for deploying more sophisticated hardware- and software-based quantization techniques. Quantizing additional tensors---such as labels and adjacency matrices---and using lower-precision formats may yield even greater speedups. Our goal was to showcase the potential of this approach to accelerate overall training, which was clearly demonstrated. As with the sampling-based method discussed earlier, the benefits of quantization are most pronounced for datasets with high edge-to-node ratios and in systems with significant communication overhead.

\section{Architectural Improvements}
Working with BlueField-2 has provided valuable insights into potential architectural improvements. This section presents a selection of the most important ones.

\subsection{High-overhead Networks}
It is important to highlight our expectations regarding larger graphs (relative to our evaluated datasets) that demonstrate good accuracy with robust sampling. In particular, those with a high edge-to-node ratio are likely to show even greater acceleration in transfer times with Neighbor Sampling offloaded to the SmartNIC. This is especially true when data is transferred across high-overhead networks such as Ethernet or InfiniBand. This is attributed to the fact that larger graphs, when sampled with significant in-layer sampling hyperparameters (e.g., [25, 10]), can substantially reduce transaction times by minimizing the number of chunks transmitted over the network without significantly compromising accuracy, as demonstrated in this paper. 

Furthermore, quantizing such graphs prior to transmission can lead to additional reductions in total training time. Specifically, when large graphs with high edge-to-node ratios---and relatively low preprocessing overhead---are quantized on the SmartNIC before being transferred over high-overhead networks, the resulting speedups are particularly pronounced. We anticipate that further optimizing preprocessing latency on the SmartNIC---by quantizing additional tensors (e.g., label and adjacency matrix tensors) and employing lower-precision formats---will yield even greater performance gains.

\subsection{NVIDIA's BlueField-2 and DOCA}
\label{subsection: NVIDIA's BlueField-2 and DOCA}
As discussed, PyG supports CPU Affinity, which binds specific cores to preprocessing tasks and allocates dedicated memory per core to enable parallel batch preprocessing. However, as shown in Figure~\ref{sampling_comparison}, this strategy degraded performance on the SmartNIC due to limited memory, since each worker replicates the data loader’s memory requirements. Consequently, we disabled CPU Affinity during training and limited experiments to small- and medium-sized graphs with few layers---a common setup, since additional layers rarely improve accuracy. Additionally, the SmartNIC’s preprocessing was 2-10x slower than the CPU due to its lower computational power. To address this, we propose increasing BlueField-2’s core count, optimizing individual cores, and expanding memory to reduce preprocessing bottlenecks.


As discussed earlier, we identified a 1 MB limitation in the DOCA-DMA buffer size, which prevented the DOCA-DMA synchronization mechanism from reaching its full performance potential. Additionally, the DOCA-DMA buffers could not be reallocated without reinitializing the DOCA process, which is time-consuming. This constraint necessitated allocating a separate local buffer and copying data from the DOCA buffer, resulting in a performance penalty. Implementing dynamic reallocation for DOCA buffers would allow devices to work directly with DOCA buffers with zero-copy. Therefore, we propose that allowing the buffer size to be increased, along with supporting the reallocation of both remote and local DOCA buffers, could significantly optimize DOCA-DMA data transfers by avoiding an extra memory copy for each chunk.

\section{Related Work}
Unlike prior work, which generally limits SmartNIC offloading to lightweight or stateless operations, LGNNIC offloads mini-batch Neighbor Sampling---a non-standard, compute-intensive operation---to the SmartNIC, achieving substantial reductions in data movement. We also offload Tensor Quantization, a more conventional optimization, to further reduce transfer overhead.

\subsection{Sampling Large Graphs}
When generating GNN embeddings, the computational graphs grow exponentially as more layers and neighbors per layer are added. Consequently, when handling large graphs, a preprocessing phase known as sampling becomes essential. Sampling, a critical component of the preprocessing phase, reduces the size of the mini-batches' computational graphs. These sampled mini-batches are then fed into the GPU for training. Various sampling categories exist, including node-wise (e.g., GraphSAGE \cite{ref:hamilton2018inductive}, VR-GC \cite{ref:vr-gcn}, PinSAGE Sampler \cite{ref:pinsage}), layer-wise (e.g., FastGCN \cite{ref:fastgcn}, AS-GCN \cite{ref:as-gcn}, LADIES \cite{ref:ladies}), and subgraph-wise (e.g., GraphSAINT \cite{ref:graphsaint}, RWT \cite{ref:rwt}).

Among node-wise methods, the most widely used approach is Neighbor Sampling, introduced by Hamilton et al. \cite{ref:hamilton2018inductive}. The Neighbor Sampling algorithm limits the number of layers in the computational trees and constrains the number of nodes per layer, thereby creating smaller computational trees for mini-batches. Within the concept of Neighbor Sampling, various strategies for selecting nodes to sample exist. These include random sampling, which may unintentionally prune important nodes, and random walk with restarts (RWR) \cite{rwr}, which ensures the inclusion of significant nodes.

The Neighbor Sampling algorithm, which is used in our system, iteratively samples nodes based on the number of GNN layers and a specified maximum number of neighbors per layer. It is important to note that even after sampling, the size of the computational graph still grows exponentially with the number of layers. To ensure stable training with minimal variance in neighbor aggregation---and to prevent a significant drop in accuracy---the average number of neighbors per node after sampling must not decrease substantially. Therefore, overly aggressive sampling of the computational graph is not feasible; it is crucial to strike a careful balance between meeting the memory constraints of GPUs or CPUs and maintaining high accuracy.

Other large-graph sampling methods exist to enable large-scale GNN training. For example, Advanced Cluster-GCN \cite{ref:Cluster-GCN} constructs small sets of node groups that are aggregated to form larger node groups, offering better resource utilization compared to Neighbor Sampling. Another example is the Simplified GNN \cite{ref:simplified_gnn} architecture, where a linear matrix is used instead of computing nonlinear activations during the forward pass, enhancing scalability at the expense of expressiveness.

\subsection{Communication Overhead}
In Farview \cite{ref:farview}, the authors alleviate data-transfer bottlenecks with a remote buffer cache that is capable of processing data streams before being transmitted over the network. It supports a few defined offloading operators, specifically for SQL queries, using RoCE v.2 RDMA transfers. 
In our research, as we also explore the potential of offloading to SmartNICs, we focus on evaluating the impact of rebalancing the entire algorithm across the compute and memory components of the training process, rather than simple SQL operations and intra-node optimizations.

The DGCL paper \cite{dgcl} also addresses communication overheads by distributing GNN training across multiple GPUs rather than across remote memory nodes. The Sequential Aggregation and Rematerialization paper \cite{SequentialAggregationAndRematerialization} addresses smart graph partitioning using aggregation and rematerialization, freeing memory across multiple processing workers during the backward pass, enabling the scaling of large GNNs.

Other works deal with GNN acceleration on a single computational node (Memory-CPU-GPU and not inter-node networks). SmartSAGE \cite{ref:lee2022smartsage}, e.g., deals with accelerating the same GraphSAGE \cite{ref:hamilton2018inductive} sampling algorithm implemented based on the same library (PyG \cite{ref:Fey/Lenssen/2019}) we use, with an in-storage architecture exploiting a smartSSD composed of NVMe SSD and an FPGA. The 
GNNear paper \cite{zhou2022gnnear} also focuses on near-memory processing to accelerate memory-intensive reduce \& update operations.

There are other works focusing on accelerating the sampling phase on CPU-FPGA systems, including pure sampling acceleration on FPGAs \cite{ref:cpu_fpga_paper1, ref:cpu_fpga_paper2} and on multi-FPGA platforms \cite{ref:cpu_fpga_paper3}.

While prior work mostly focuses on quantizing activations and weights \cite{ref:improving_speed_ofnn, ref:binarized, ref:deep_compression, ref:dcn_quant, ref:gnn_quantization1, ref:gnn_quantization2, ref:gnn_quantization3}, our approach quantizes data before transmission and dequantizes it at the receiver prior to training.

\section{Conclusions and Next Steps}
We introduced LGNNIC, a novel architecture for accelerating large-scale GNN training using SmartNICs. LGNNIC offloads Neighbor Sampling and Quantization to the SmartNIC, reducing data transfer overhead and improving training time. To evaluate this architecture on our PoC system, we developed a low-overhead DOCA-DMA synchronization mechanism for efficient buffer management and compared it to a high-overhead socket-based benchmark.

Offloading Neighbor Sampling consistently yielded end-to-end training speedups, particularly in transaction time, while quantization reduced transfer time, especially under network congestion. Even modest FP32-to-FP16 quantization provided substantial benefits, and additional gains are possible by quantizing more tensors or using lower precisions. Performing these operations on the SmartNIC avoids unnecessary data movements between remote and host, making it a natural and efficient location for this task. The results across three datasets, multiple sampling configurations, and two communication mechanisms demonstrate the feasibility of LGNNIC's per-node SmartNIC offloading approach.

Evaluating LGNNIC in a full multi-node deployment, including cross-partition sampling, inter-node coordination, routing, and scalability, is left to future work. Moreover, we plan to explore peer-to-peer DOCA-RDMA over Ethernet for higher-latency distributed setups and for zero-copy transfers from remote SmartNIC DRAM to the local GPU. We will profile the system, identify bottlenecks, and assess additional acceleration on BlueField-2, as well as evaluate more advanced BlueField-3 and BlueField-4 devices, which offer greater memory capacity and bandwidth and may further reduce preprocessing and communication overhead, thereby improving overall training performance.
\bibliographystyle{IEEEtran}
\bibliography{sampling}


\appendix

\section{Appendix}
\subsection{Neighbor Sampling With PyG}
\label{subsection: appendix: Neighbor Sampling With PyG}

In PyG's mini-batch training with Neighbor Sampling, each epoch iterates over the \texttt{NeighborLoader} class (a subclass of PyTorch’s \texttt{DataLoader}), typically instantiated as \texttt{train\_loader}, to generate sampled mini-batches. The tensors from these batches---including node indices, features, and labels---are transferred to the GPU, where the remaining training steps, such as the forward pass, loss computation, backpropagation, and parameter updates, are executed.

\begin{algorithm}
\caption{PyG mini-batch training with Neighbor Sampling} \label{alg:simple_training}
\begin{algorithmic}[1]
    \State \textbf{Initialize:} Dataset
    \State \textbf{Initialize:} Train\_Loader Configuration
    \State \textbf{Initialize:} Model
    \State \textbf{Initialize:} Optimizer
    
    \For{$\text{epoch} = 1$ \textbf{to} $\text{Epoch Number}$}
        \For{\textbf{each} mini-batch \textbf{in} Train\_Loader}
            \State \textbf{Transfer:} Mini-Batch Data to GPU
            \State \textbf{Perform:} Forward Pass Through The Model
            \State \textbf{Calculate:} Loss Using Predicted and Actual Labels
            \State \textbf{Calculate:} Gradients via Backpropagation
            \State \textbf{Update:} Model Parameters Using Optimizer
        \EndFor
    \EndFor
\end{algorithmic}
\end{algorithm}

\subsection{Overall Training Speedups through Offloading Tensor Quantization}
\label{subsection: appendix: Overall Training Speedup through Offloading Tensor Quantization}
In Table~\ref{tab:acceleration_results_quantization}, we compare FP32 and FP16 precision when quantizing the feature tensor to highlight the impact of quantization on total transfer time and, consequently, total training (execution) time. Here, in Table~\ref{tab:acceleration_results_quantization_appendix}, we present the detailed results of the measured average total transaction time, average total execution time, and the resulting speedups that were derived from them. The measurements were conducted across all mechanisms (DOCA-DMA and Sockets), datasets, and sampling hyperparameters (with common hyperparameters highlighted in yellow). 

Notably, in settings where transaction time dominates overall training time---such as with the high-overhead Sockets mechanism---the execution speedup closely tracks the transfer speedup. This effect is particularly pronounced in the Reddit dataset, which has the highest edge-to-node ratio among those evaluated. The high edge-to-node ratio also explains why Reddit exhibits the highest total training time among the evaluated datasets with Sockets, while preprocessing time remains moderate. In contrast, when transaction time constitutes a smaller fraction of total training time---as is typically the case with the low-overhead DOCA-DMA mechanism, where preprocessing is also a significant bottleneck---the execution and transfer speedups diverge. Since DOCA-DMA reduces communication overhead substantially, the impact of quantization is less pronounced compared to Sockets, where communication overhead remains the primary bottleneck.

\begin{table*}[h!]
\centering
\begin{subtable}[t]{0.5\linewidth}
\centering
\scriptsize
\setlength{\tabcolsep}{3pt}
\begin{tabular}{|>{\columncolor{gray!20}}c|>{\columncolor{gray!20}}c|c|c|>{\columncolor{gray!20}}c|c|c|}
\hline
\rowcolor{gray!20}
& \makecell{\textbf{Num} \\ \textbf{Neighbors}} 
& \makecell{\textbf{Tot. Trans. Time} \\ \textbf{[s] (FP32 / FP16)}} 
& \makecell{\textbf{Speed} \\ \textbf{-up}} &
& \makecell{\textbf{Tot. Exec. Time} \\ \textbf{[s] (FP32 / FP16)}}
& \makecell{\textbf{Speed} \\ \textbf{-up}}  \\
\hline
& \([-1, 256]\) & 315.52 / 192.87 & 1.64 && 325.84 / 201.43 & 1.62 \\
\textbf{Sockets} & \([25, 10]\) & \cellcolor{yellow}36.84 / 10.11 & \cellcolor{yellow}3.64 && \cellcolor{yellow}37.19 / 10.45 & \cellcolor{yellow}3.56 \\
& \([10, 10]\) & 14.80 / 4.11 & 3.60 && 15.00 / 4.31 & 3.48 \\
& \([-1]\) & 60.75 / 16.65 & 3.65 && 61.10 / 17.00 & 3.59 \\
\hline
& \([-1, 256]\) & 1.73 / 1.39 & 1.24 && 11.54 / 11.18 & 1.03 \\
\makecell{\textbf{DOCA} \\ \textbf{-DMA}} & \([25, 10]\) & \cellcolor{yellow}0.34 / 0.25 & \cellcolor{yellow}1.36 && \cellcolor{yellow}0.66 / 0.55 & \cellcolor{yellow}1.20 \\
& \([10, 10]\) & 0.22 / 0.17 & 1.24 && 0.40 / 0.35 & 1.15 \\
& \([-1]\) & 0.44 / 0.33 & 1.37 && 0.76 / 0.61 & 1.24 \\
\hline
\end{tabular}
\caption{Reddit}
\label{table:reddit_times}
\end{subtable}
\hfill
\begin{subtable}[t]{0.5\linewidth}
\centering
\scriptsize
\setlength{\tabcolsep}{3pt}
\begin{tabular}{|>{\columncolor{gray!20}}c|>{\columncolor{gray!20}}c|c|c|>{\columncolor{gray!20}}c|c|c|}
\hline
\rowcolor{gray!20}
& \makecell{\textbf{Num} \\ \textbf{Neighbors}} 
& \makecell{\textbf{Tot. Trans. Time} \\ \textbf{[s] (FP32 / FP16)}} 
& \makecell{\textbf{Speed} \\ \textbf{-up}} &
& \makecell{\textbf{Tot. Exec. Time} \\ \textbf{[s] (FP32 / FP16)}}
& \makecell{\textbf{Speed} \\ \textbf{-up}} \\
\hline
& \([-1, -1]\) & 305.76 / 135.99 & 2.25 && 318.23 / 147.03 & 2.16 \\
\textbf{Sockets} & \([15, 10, 5]\) & \cellcolor{yellow}15.65 / 4.61 & \cellcolor{yellow}3.39 && \cellcolor{yellow}17.22 / 6.07 & \cellcolor{yellow}2.84 \\
& \([25, 10]\) & 3.62 / 1.00 & 3.63 && 4.24 / 1.56 & 2.71 \\
& \([-1]\) & 2.59 / 0.57 & 4.53 && 2.97 / 0.90 & 3.31 \\
\hline
& \([-1, -1]\) & 1.59 / 1.28 & 1.24 && 13.81 / 13.60 & 1.02 \\
\makecell{\textbf{DOCA} \\ \textbf{-DMA}} & \([15, 10, 5]\) & \cellcolor{yellow}0.25 / 0.17 & \cellcolor{yellow}1.41 && \cellcolor{yellow}1.87 / 1.79 & \cellcolor{yellow}1.04 \\
& \([25, 10]\) & 0.10 / 0.09 & 1.18 && 0.75 / 0.73 & 1.03 \\
& \([-1]\) & 0.09 / 0.04 & 2.02&& 0.48 / 0.43 & 1.11 \\
\hline
\end{tabular}
\caption{OGBN-Products}
\label{table:products_times}
\end{subtable}
\hfill
\begin{subtable}[t]{0.5\linewidth}
\centering
\scriptsize
\setlength{\tabcolsep}{3pt}
\begin{tabular}{|>{\columncolor{gray!20}}c|>{\columncolor{gray!20}}c|c|c|>{\columncolor{gray!20}}c|c|c|}
\hline
\rowcolor{gray!20}
& \makecell{\textbf{Num} \\ \textbf{Neighbors}} 
& \makecell{\textbf{Tot. Trans. Time} \\ \textbf{[s] (FP32 / FP16)}} 
& \makecell{\textbf{Speed} \\ \textbf{-up}} &
& \makecell{\textbf{Tot. Exec. Time} \\ \textbf{[s] (FP32 / FP16)}}
& \makecell{\textbf{Speed} \\ \textbf{-up}} \\
\hline
& \([-1, -1]\) & 87.02 / 28.75 & 3.03 && 176.76 / 59.96 & 2.95 \\
\textbf{Sockets} & \([25, 10]\) & 1.65 / 0.76 & 2.16 && 3.62 / 1.80 & 2.01 \\
& \([10, 10]\) & \cellcolor{yellow}1.18 / 0.60 & \cellcolor{yellow}1.96 && \cellcolor{yellow}2.62 / 1.43 & \cellcolor{yellow}1.83 \\
& \([-1]\) & 0.18 / 0.15 & 1.20 && 0.44 / 0.36 & 1.20 \\
\hline
& \([-1, -1]\) & 0.91 / 0.69 & 1.33 && 4.09 / 3.62 & 1.13 \\
\makecell{\textbf{DOCA} \\ \textbf{-DMA}} & \([25, 10]\) & 0.12 / 0.06 & 1.93 && 0.55 / 0.44 & 1.25 \\
& \([10, 10]\) & \cellcolor{yellow}0.10 / 0.05 & \cellcolor{yellow}1.96 && \cellcolor{yellow}0.47 / 0.36 & \cellcolor{yellow}1.31 \\
& \([-1]\) & 0.02 / 0.017 & 1.12 && 0.10 / 0.10 & 1.05 \\
\hline
\end{tabular}
\caption{OGBN-MAG}
\label{table:mag_times}
\end{subtable}

\caption{Comparison of FP32 and FP16 precision: average total transaction time, average total execution time, and corresponding speedups across all mechanisms, datasets, and sampling hyperparameters.}
\label{tab:acceleration_results_quantization_appendix}
\end{table*}

\end{document}